\documentclass[aps,prb,twocolumn,preprintnumbers,amsmath,amssymb,aps,superscriptaddress,floatfix,byrevtex,longbibliography,latexsym]{revtex4-2}
\usepackage{graphicx}
\usepackage{dcolumn}
\usepackage{bm}
\usepackage{appendix}
\usepackage{wrapfig}

\usepackage{physics}

\usepackage[colorlinks=true, linkcolor=blue, citecolor=blue, urlcolor=blue]{hyperref}

\font\scripti=cmmi7
\font\scriptscripti=cmmi5
\def\sib#1{\setbox0 = \hbox{\scripti #1}
  \kern-.02em\copy0\kern-\wd0
  \kern.04em\box0} 
\def\ssib#1{\setbox0 = \hbox{\scriptscripti #1}
  \kern-.02em\copy0\kern-\wd0
  \kern.04em\box0} 
\font\tenib=cmmib10 
\skewchar\tenib='177 \skewchar\tenib='177 \skewchar\tenib='177
\def\pbold#1{\setbox0 = \hbox{$ #1 $}
  \kern-.022em\copy0\kern-\wd0
  \kern.011em\copy0\kern-\wd0
  \kern.011em\copy0\kern-\wd0
  \kern.011em\copy0\kern-\wd0
  \kern.011em\box0} 

\newcommand{\beginsupplement}{%
        \setcounter{table}{0}
        \renewcommand{\thetable}{S\arabic{table}}%
        \setcounter{figure}{0}
        \renewcommand{\thefigure}{S\arabic{figure}}%
      \setcounter{equation}{0}
        \renewcommand{\theequation}{S.\arabic{equation}}%

     }

\newcommand{\red}[1]{{#1}}

\usepackage{graphicx}
\usepackage{dcolumn}
\usepackage{bm}
\usepackage{color}

\def\up{\uparrow}
\def\dwn{\downarrow}

\def\lesssim{\ \raise.3ex\hbox{$<$}\kern-0.8em\lower.7ex\hbox{$\sim$}\ }
\def\gesim{\ \raise.3ex\hbox{$>$}\kern-0.8em\lower.7ex\hbox{$\sim$}\ }

\begin{document}

\title{Emergent Fano-Feshbach resonance in two-band superconductors with an incipient quasi-flat band: Enhanced critical temperature evading particle-hole fluctuations}

\author{Hiroyuki Tajima}
\affiliation{Department of Physics, Graduate School of Science, 
The University of Tokyo, Hongo, Tokyo 113-0033, Japan}
\affiliation{RIKEN Nishina Center, Wako 351-0198, Japan}
\author{Hideo Aoki}
\affiliation{Department of Physics, Graduate School of Science, The University of Tokyo, Hongo, Tokyo 113-0033, Japan}
\author{Andrea Perali}
\affiliation{School of Pharmacy, Physics Unit, Universit\'{a} di Camerino, 62032 Camerino (MC), Italy}
\author{Antonio Bianconi}
\affiliation{Rome International Center for Materials Science Superstripes RICMASS, 
00185 Roma, Italy}
\affiliation{Institute of Crystallography, Italian National Research Council, IC-CNR, 
00015 Roma, Italy}

\date{\today}
\begin{abstract} 
In superconductivity, a surge of interests in enhancing $T_{\rm c}$ is ever mounting, where a
recent focus is toward multi-band superconductivity.  
In $T_{\rm c}$ enhancements specific to two-band cases, 
especially around the Bardeen-Cooper-Schrieffer (BCS) to Bose-Einstein condensate (BEC) crossover considered here, we have
to be careful about how quantum fluctuations
affect the many-body states, i.e., particle-hole fluctuations 
suppressing the pairing for attractive interactions.  Here we explore how to circumvent the suppression by examining 
multichannel pairing interactions in two-band systems.  With the Gor'kov-Melik-Barkhudarov (GMB) 
formalism for particle-hole fluctuations in a continuous space, we look into 
the case of a deep dispersive band accompanied by an incipient heavy-mass (i.e., quasi-flat) band.  We find that, while the GMB corrections usually 
suppress $T_{\rm c}$ significantly, 
this in fact competes with the enhanced pairing 
arising from the heavy band, with the trade-off 
leading to a {\it peaked} structure in $T_{\rm c}$ against the band-mass ratio when the heavy band is incipient.
The system then plunges into a strong-coupling regime with the GMB screening vastly suppressed. This occurs prominently when the chemical potential approaches the bound state lurking just below the 
heavy band, which can be viewed as a Fano-Feshbach resonance, with its width governed by the pair-exchange interaction.
The diagrammatic structure comprising particle-particle and particle-hole channels is heavily entangled, so that the emergent Fano-Feshbach resonance 
dominates all the channels, suggesting a universal feature in multiband superconductivity and superfluidity.
\end{abstract}
\maketitle
\noindent{\it Introduction}---
Multi-band electronic systems and their multi-component superconducting phases can harbor 
novel quantum effects.  
Superconductivity and its microscopic theory initiated by Bardeen, Cooper and Schrieffer (BCS) give a conceptual impact on various research fields that encompass nuclear and particle physics as well~\cite{PhysRev.110.936,PhysRev.122.345,PhysRev.124.246}.
Moreover, the discoveries of high-$T_{\rm c}$ superconductors, such as cuprates~\cite{bednorz1986possible} and iron pnictides~\cite{kamihara2008iron}, have ignited renewed interests toward the realization of higher-temperature superconductivity.

Crucial factors for amplification of superconductivity are mainly two-fold: the interparticle interaction and the electronic band structure.  For an attractive interaction, the question is designing the ways to enhance the magnitude of the interaction for one-band cases.  A pivotal factor then 
is the crossover from the BCS regime with loosely-bound Cooper pairs to the Bose-Einstein condensation (BEC) regime with 
tightly-bound pairs when the strength of the attraction is 
increased and/or the carrier density is reduced~\cite{PhysRev.186.456,leggett2008diatomic,nozieres1985bose,PhysRevLett.71.3202}.
While it is difficult to control the interaction in situ in condensed matters,
the BCS-BEC crossover was realized about two decades ago in ultracold Fermi gases near the Fano-Feshbach resonance~\cite{RevModPhys.82.1225,PhysRevLett.92.040403,PhysRevLett.92.120403,PhysRevLett.92.203201}.
Recently, the realization of solid-state systems in 
the BCS-BEC crossover regime has also been reported in FeSe superconductors~\cite{lubashevsky2012shallow,kasahara2014field,doi:10.1126/sciadv.1602372,doi:10.1126/sciadv.abb9052,mizukami2023unusual},
Li$_x$ZrNCl~\cite{PhysRevB.98.064512,nakagawa2021gate},
and organic superconductors~\cite{PhysRevX.12.011016}
by tuning carrier densities.

If we go over to 
multi-band superconductors and superfluids, 
the increased degrees of freedom can host diverse  quantum phenomena~\cite{Milosevic_2015}. For example, a multi-band configuration with shallow and deep bands plays a crucial role typically in FeSe~\cite{PhysRevB.93.174516}.
A remarkable feature of the multi-band BCS-BEC crossover is a reduction of pairing fluctuations in the strong-coupling regime~\cite{PhysRevB.100.064510,PhysRevB.99.180503,PhysRevB.102.220504,PhysRevLett.125.217003} which tends to suppress the superconducting critical temperature $T_{\rm c}$. This screening effect is
consistent with the observation of missing pseudogap in FeSe~\cite{PhysRevLett.122.077001}, whereas the pseudogap induced by pairing fluctuations is expected in the single-band BCS-BEC crossover~\cite{perali2002,chen2005bcs,mueller2017review,strinati2018bcs,OHASHI2020103739}.
Regarding the realization of strong-coupling systems, geometrical quantum confinement in the form of slabs or stripes causes interference between wavefunctions associated to different subbands, inducing superconducting shape resonances when the chemical potential is close to one of the subband bottom~\cite{PhysRevB.82.184528}.
Moreover, an interband pair-exchange coupling in two-band systems leads to a kind of the Suhl-Kondo mechanism~\cite{PhysRevLett.3.552,10.1143/PTP.29.1} which modifies the effective attraction in each band~\cite{PhysRevB.88.014509,PhysRevB.106.L180506,PhysRevResearch.4.013032}, so that this can be evoked for realizing the BCS-BEC crossover.
Unconventional phase transitions
have also been reported even in simple two-band 
models without complicated band structures nor  impurities~\cite{PhysRevB.74.144517,PhysRevB.84.094515,PhysRevB.85.134514,PhysRevLett.108.207002,PhysRevB.100.104528,PhysRevB.107.184501}. 
Such multi-band characters may be further enhanced when the effective mass in the second band is heavy (flat or quasi-flat band)~\cite{aoki2020theoretical}.
Recently, a resonant enhancement of $T_{\rm c}$ in a multi-band system near a topological Lifshitz transition has also been studied in spin-orbit-coupled artificial superlattices~\cite{PhysRevB.103.024523,10.1063/5.0123429,condmat8030078}.

The multi-band BCS-BEC crossover has been studied intensively, but an important point about fluctuations is still unclear.  Namely, in single-band models, 
the particle-hole fluctuations for attractive interactions, as formulated by Gork'ov-Melik-Barkhudarov (GMB)~\cite{gor1961contribution}, significantly reduce $T_{\rm c}$.  
Quantitatively, the GMB correction is known to 
reduce $T_{\rm c}$ in single-band systems by a factor $(4e)^{1/3}\simeq 2.2$ in the weak-coupling (BCS) limit~\cite{PhysRevB.97.014528}.  Thus an imperative question is to find out  
how the GMB correction arises in two-band systems.  
This becomes crucial, in our view, when the second band is \textit{incipient}, where the chemical potential $\mu$ is close to the bottom of the second band and the band 
starts to be occupied.  Intuitively,  
this situation is expected to strongly affect the interaction via the pair-exchange coupling. 
 
Recently, the evolution of $T_{\rm c}$ along the BCS-BEC crossover has been experimentally detected in a single-band ultracold system of atomic fermions~\cite{PhysRevLett.130.203401}.  By comparing with the theoretical prediction of the GMB, which was 
originally devised for the BCS regime but then extended to the BCS-BEC crossover in Ref.~\cite{PhysRevB.97.014528}, the existence of the GMB correction and its evolution has indeed been experimentally confirmed, after 60 years of the pioneering GMB paper~\cite{gor1961contribution}. 
Thus the question of how the GMB correction on $T_{\rm c}$ discussed in Ref.~\cite{PhysRevB.97.014528} will behave in two-band models is of both fundamental and practical importance.

Motivated by these backgrounds, the present work theoretically explores the GMB screening effects on $T_{\rm c}$ in a two-band system consisting of a dispersive (light-mass) band and a quasi-flat (heavy-mass) band with intraband attractive interactions accompanied by interband pair-exchange couplings. 
In particular, we focus on the situation where the heavy band is {\it incipient} (with the chemical potential close to the bottom of the second band) to fathom how the heavy band can dominate the dispersive band.
For that, we have extended the GMB approach to two-band systems in terms of the simplified diagrammatic approach developed in Ref.~\cite{PhysRevA.79.053636}.  
In particular, 
the different effective masses of dispersive and heavy bands are considered here,
in contrast to Ref.~\cite{PhysRevB.93.174516} where 
only equal-mass two bands were considered and hence the effects of the incipient quasi-flat band were unraveled.

\begin{figure}[t]
\begin{center}
\includegraphics[width=8.6cm]{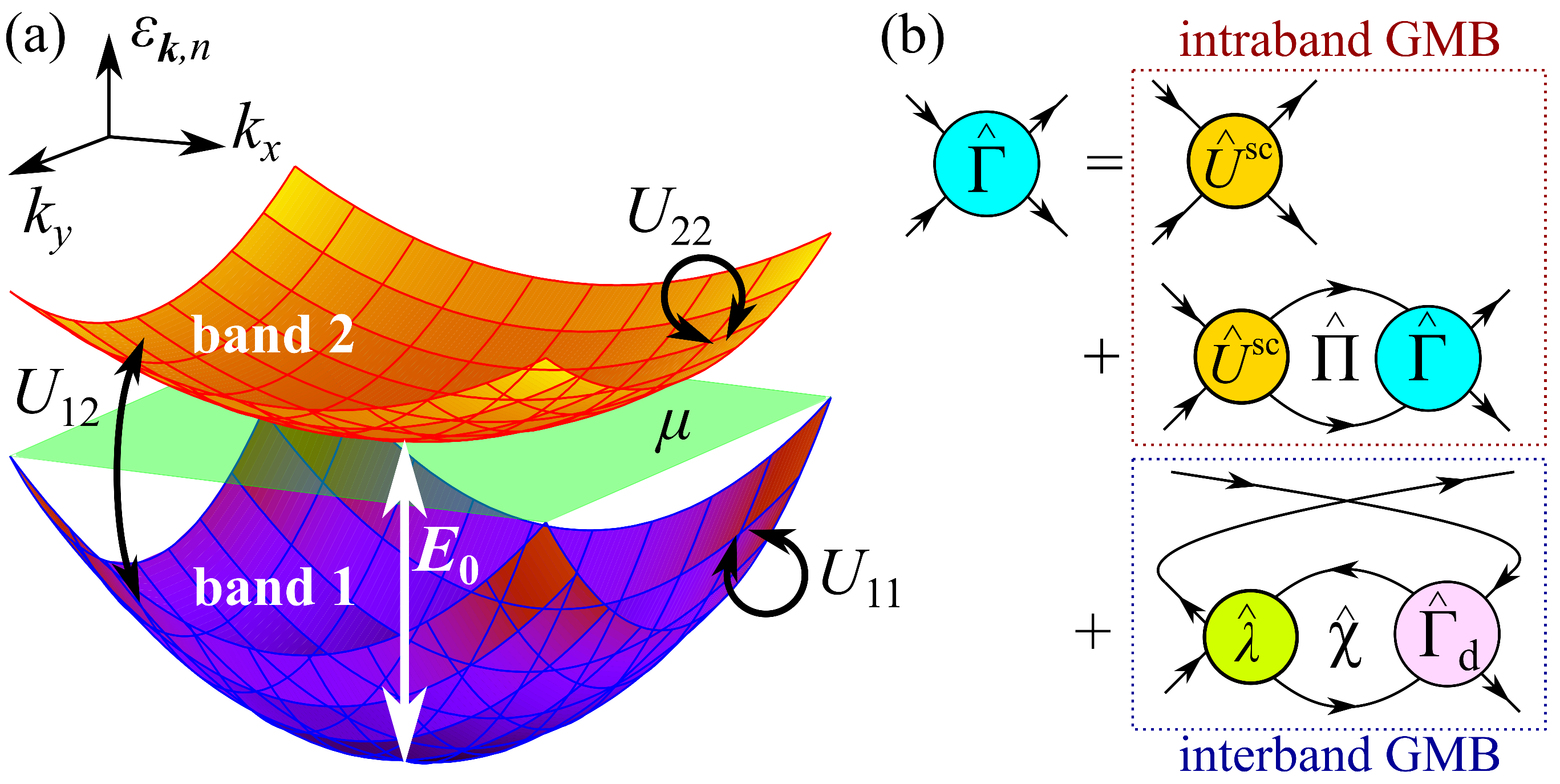}
\end{center}
\caption{(a) The band structure of the two-band model considered here with a light-mass 
1st band and a heavy-mass 2nd band with intraband ($U_{11}$, $U_{22}$) and interband ($U_{12}=U_{21}$) couplings, and  the band offset $E_0$.
The chemical potential $\mu$ is set to be close to the bottom of the 2nd band.
(b) Diagrammatic representation of the many-body $T$-matrix $\hat{\Gamma}$ in the GMB formalism that comprises the intraband interaction through the screened coupling $\hat{U}^{\rm sc}$ (encircled in red), and the 
GMB correction for the interband pair-exchange interaction through the pair-exchange-induced coupling $\hat{\lambda}$ and the diagonal component $\hat{\Gamma}_{\rm d}$ (encircled in blue). 
}
\label{fig:1}
\end{figure}

\noindent{\it Two-band system composed of dispersive and heavy bands}---
We consider a two-band model in continuum in three dimensions described by the Hamiltonian
\begin{align}
\label{eq1}
H=&\sum_{\bm{k},\sigma,n}\xi_{\bm{k},n}c_{\bm{k},\sigma,n}^{\dag}c_{\bm{k},\sigma,n} 
+
\sum_{\bm{k},\bm{k}',\bm{q},n,n'}U_{nn'}b_{\bm{k},\bm{q},n}^{\dag}b_{\bm{k}',\bm{q},n'},
\end{align}
where $c_{\bm{k},\sigma,n}^{\dag}$ creates a fermion with momentum $\bm{k}$ and spin $\sigma=\up,\dwn$ in band $n(=1,2)$, and $b_{\bm{k},\bm{q},n}^{\dag} \equiv c_{\bm{k}+\bm{q}/2,\up,n}^{\dag}c_{-\bm{k}+\bm{q}/2,\dwn,n}^{\dag}$
is a pair-creation operator.
$\xi_{\bm{k},n}=\varepsilon_{\bm{k},n}-\mu+E_0\delta_{n,2}$ is the kinetic energy in band $n$ measured from the chemical potential $\mu$, where $\varepsilon_{\bm{k},n}=k^2/2m_n$ ($m_n$ is the effective mass of band $n$) and $E_0$
is the band offset between the two bands [see Fig.~\ref{fig:1}(a)].
In this work, we assume that the upper band ($n=2$) has a heavier effective mass, $m_2\geq m_1$.

To characterize the intraband interaction strength $U_{nn}$, 
we use a scattering length $a_{nn}$ given by
$\frac{m_n}{4\pi a_{nn}}
=
U_{nn}^{-1}
+\frac{m_n\Lambda}{2\pi^2}$
for $n=1,2$, where $\Lambda$ is the momentum cutoff that is needed in continuum models~\cite{PhysRevB.74.144517}.  
We can roughly translate $\Lambda$ as 
the band width in lattice models.  
For simplicity, we assume that the intraband interaction is independent of the band index, $U_{22}=U_{11}$.  
The coupling $U_{11}$ within the dispersive band is kept weak in such a way that the corresponding scattering length is negative, $k_0a_{11}=-1.0$ here.
The interband pair-exchange couplings are $U_{12}$ and $U_{21} 
(=U_{21}$ for the Hermiticity of $H$).  It is convenient to introduce 
a dimensionless coupling as
$\tilde{U}_{12}\equiv U_{12}\frac{\sqrt{m_1m_2}k_0}{2\pi^2} = \tilde{U}_{21}$,
where we have introduced a momentum scale $k_0\equiv\sqrt{2m_1 E_0}$.

\noindent
\textit{Many-body $T$-matrix with particle-hole fluctuations}---
Let us now present the equation for $T_{\rm c}$ with the GMB screening effect in the present two-band system based on the diagrammatic approach.
As displayed in Fig.~\ref{fig:1}(b), the many-body $T$-matrix 
$\hat{\Gamma}$ in the $2\times 2$ matrix representation for band indices reads
\begin{align}
\label{eq:gamma}
    \hat{\Gamma}(q)
    &=
    \hat{U}^{\rm sc}-
    \hat{U}^{\rm sc}\hat{\Pi}(q)\hat{\Gamma}(q)
    -\hat{\lambda}(q)\langle \hat{\chi}\rangle
    \hat{\Gamma}_{\rm d}(q),
\end{align}
where $q=(\bm{q},i\nu_\ell)$ is the four-momentum index with boson Matsubara frequency $\nu_\ell=2\pi\ell T$ $(\ell\in \mathbb{Z})$, and
\begin{align}
\label{eq:usc}
    \hat{U}^{\rm sc}=\left(
    \begin{array}{cc}
      U_{11}^{\rm sc}   & U_{12}  \\
      U_{21}   & U_{22}^{\rm sc} 
    \end{array}
    \right)
\end{align}
is the coupling constant matrix.  Its diagonal components involve the GMB screening for $U_{11}$ and $U_{22}$ as
 $U_{nn}^{\rm sc}=U_{nn}/(1+U_{nn}\langle \chi_{nn}\rangle)$ with the averaged particle-hole bubble $\langle\chi_{nn}\rangle$~\cite{PhysRevA.79.053636}.
 Here we have simplified the framework following Ref.~\cite{PhysRevB.97.014528}, which should 
 be qualitatively valid, as indicated by 
 $T_{\rm c}$ in Ref.~\cite{PhysRevA.79.053636} being similar to Ref.~\cite{PhysRevB.97.014528} across the BCS-BEC crossover.  
 For $\mu-E_0\delta_{n,2}>0$, we obtain
\begin{align}
\label{eq:phb}
\left<\chi_{nn}\right>=\frac{m_n}{4\pi^2}{}\int_{-1}^{1}ds\int_{0}^{\infty}\frac{kdk}{q_n}f(\xi_{\bm{k},n}){\rm ln}\left|\frac{q_n-2k}{q_n+2k}\right|,
\end{align}
where we have defined $q_n\equiv \sqrt{2m_n(\mu-E_0\delta_{n,2})(1+s)}$ and the Fermi distribution function $f(\xi_{\bm{k},n})=(e^{\xi_{\bm{k},n}/T}+1)^{-1}$.
When $\mu-E_0\delta_{n,2}<0$ where the Fermi surface is absent for band $n$, we get
$\left<\chi_{nn}\right>=-\frac{m_n}{2\pi^2}\int_{0}^{\infty}dkf(\xi_{\bm{k},n})$.
\red{This treatment reflects an aspect that the particle-hole bubble is strongly suppressed in the BEC regime, where the chemical potential strongly deviates from the Fermi energy that in the weak-coupling limit is given by $E_{{\rm F},n}=(3\pi^2\rho_{n})^{2/3}(2m_n)^{-1}$ for a given number density $\rho_n$~\cite{PhysRevB.99.180503}
and can become negative~\cite{PhysRevB.97.014528,PhysRevA.79.053636}, leading to a progressively exponential suppression of the particle-hole bubble of the GMB correction.}
In Eq.(\ref{eq:gamma}), 
$\hat{\Pi}(q)={\rm diag}[\Pi_{11}(q),\Pi_{22}(q)]$ is the particle-particle bubble with
\begin{align}
    \Pi_{nn}(q)=-\sum_{\bm{k}}\frac{1-f(\xi_{\bm{k}+\bm{q},n})-f(\xi_{-\bm{k},n})}{i\nu_\ell-\xi_{\bm{k}+\bm{q},n}-\xi_{-\bm{k},n}}.
\end{align}
The very last term of Eq.~\eqref{eq:gamma} (Fig.1(b), bottom line) represents the GMB correction 
(see Fig.1(b)), which consists of the pair-exchange induced coupling $\hat{\lambda}(q)={\rm diag}\left(-U_{12}U_{21}\Pi_{22}(q), -U_{12}U_{21}\Pi_{11}(q)\right)$, along 
with the particle-hole bubble $\langle\hat{\chi}\rangle \equiv {\rm diag}(\langle\chi_{11}\rangle,\langle\chi_{22}\rangle)$ and the diagonal component of the $T$-matrix, $\Gamma_{\rm d}(q)={\rm diag}\left(\Gamma_{11}(q),\Gamma_{22}(q)\right)$, 
so that 
particle-particle and particle-hole channels are heavily entangled.
Based on the Thouless criterion~\cite{THOULESS1960553}, 
$T_{\rm c}$ is obtained where $[\Gamma_{ij}(q=0)]^{-1}=0$ is achieved~\cite{PhysRevB.99.180503,PhysRevB.102.220504}.
For details about the formalism, see Supplement~\cite{Supplement}.

\begin{figure}[t]
    \centering
    \includegraphics[width=8.6cm]{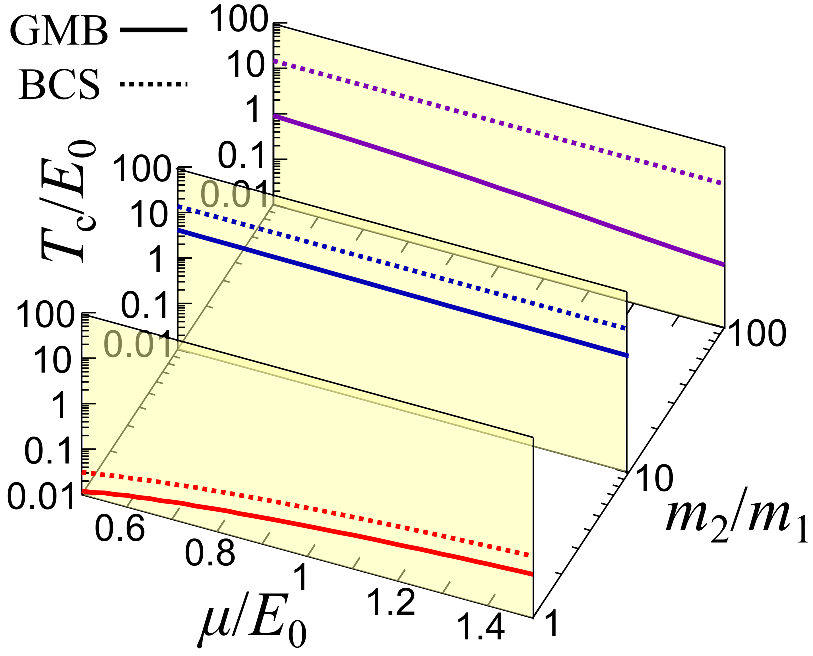}
    \caption{Superconducting critical temperature $T_{\rm c}$ against the effective mass ratio $m_2/m_1$ and chemical potential $\mu/E_0$ at $\tilde{U}_{12}=10^{-3}$ and $\Lambda/k_0=10$. The solid and dotted curves show the GMB and BCS results, respectively.}
    \label{fig:2}
\end{figure}

\noindent\textit{Interplay between pairing and particle-hole fluctuations}---
 Let us now present the numerical result for the superconducting critical temperature $T_{\rm c}$ incorporating the GMB correction in Fig.~\ref{fig:2}, where the pair-exchange coupling is set to be $\tilde{U}_{12}=10^{-3}$
 \red{(we frequently employ this value to discuss the effect of heavy mass $m_2$ in the main text.
 For different $\tilde{U}_{12}$,
 see Supplement~\cite{Supplement})}. 
 For comparison, the BCS result without the GMB correction is also displayed. 
Large enhancements in $T_{\rm c}$ can be found for 
large $m_2/m_1=10, 100$, particularly in BCS but also for GMB.
In the limit of $m_2/m_1\rightarrow \infty$, 
the Thouless criterion
without the GMB correction
simplifies to~\cite{Supplement}
\begin{align}
\label{eq:gapeq_largemass}
1+\frac{U_{22}^{\rm eff}\Lambda^3}{6\pi^2}\mathcal{F}(E_0-\mu)=0,
\end{align}
where 
    $U_{22}^{\rm eff} \equiv U_{22}
    -U_{12}\Pi_{11}(0)U_{21}/[1+U_{11}\Pi_{11}(0)]
    $,
and we have defined $\mathcal{F}(x)\equiv \frac{\tanh(x/2T_{\rm c})}{2x}$, which exhibits a maximum at $x=0$ (i.e., $\mu=E_0$).
In this limit, Eq.~\eqref{eq:gapeq_largemass} can be easily satisfied around $\mu=E_0$ even for small $U_{11}$ and $U_{22}$ for sufficiently large $\Lambda$.
While this fact is reminiscent of the enhanced pairing near the Lifshitz transition around a van Hove singularity,
the BCS result with larger $m_2/m_1$ in Fig.~\ref{fig:2} shows a weak $\mu$ dependence because the width of $\mathcal{F}(E_0-\mu)$ is $\sim T_{\rm c}/E_0 (\simeq 14$ here).
We note that the strong enhancement of $T_{\rm c}/E_0$ is associated with the cutoff-dependent effective interaction $U_{22}^{\rm eff}\Lambda^3$ in Eq.~\eqref{eq:gapeq_largemass}. In other words, $T_{\rm c}$ depends on how far the quasi-flat dispersion extends in the momentum space in band 2.

If we turn to the GMB result (solid lines in Fig.~\ref{fig:2}),
we find that the GMB correction significantly reduces $T_{\rm c}$ 
from the BCS result, particularly for large $m_2/m_1$.
This comes from the particle-hole bubble $\langle \chi_{22}\rangle$, 
which blows up for large $m_2/m_1$.
For $\mu-E_0\leq 0$, we have an expression 
\begin{align}
\label{eq:fugacity}
    \langle\chi_{22}\rangle=\frac{m_2\sqrt{2\pi m_2T}}{4\pi^2}{\rm Li}_{1/2}(-z),
\end{align}
where $z=e^{(\mu-E_0)/T_{\rm c}}$ is an effective fugacity and ${\rm Li}_{s}(x)$ is the polylogarithm.
Specifically, for $\mu\rightarrow E_0$ (i.e., $z\rightarrow 1$),
we have $\langle \chi_{22}\rangle=-\frac{m_2\sqrt{2\pi m_2T}}{4\pi^2}(1-\sqrt{2})\zeta(1/2)$ with the Riemann zeta function $\zeta(1/2)\simeq -1.46$.
This leads to a divergent behavior of $\langle \chi_{22}\rangle\propto m_2^{3/2}$ for $m_2\rightarrow\infty$.
Such a tendency persists for $\mu>E_0$ as seen in Fig.~\ref{fig:2}.

\begin{figure}[t]
    \centering
    \includegraphics[width=8.6cm]{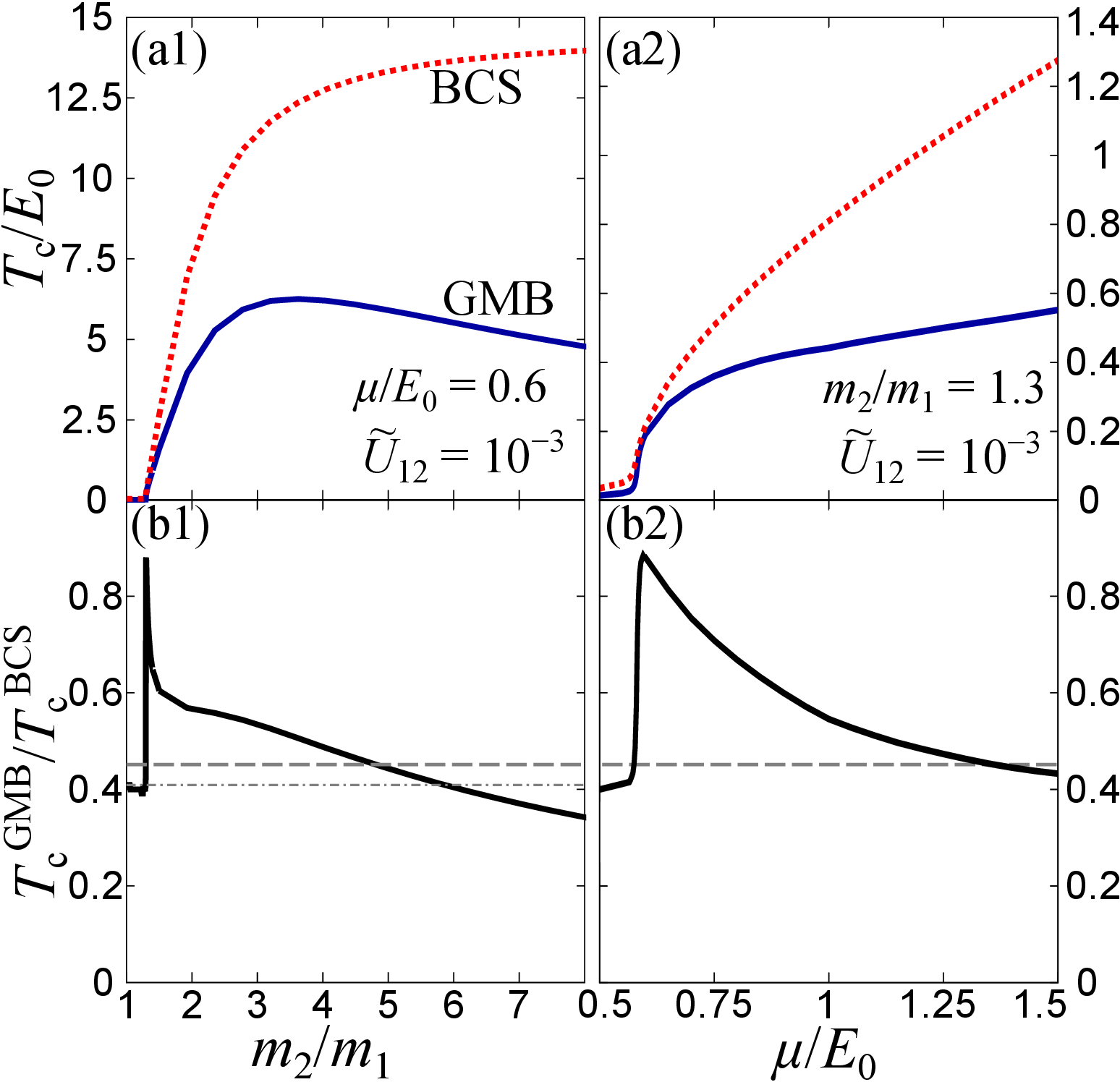}
    \caption{(a1) Superconducting critical temperatures $T_{\rm c}$ as a function of the effective mass ratio $m_2/m_1$ at $\mu/E_0=0.6$. Panel (a2) shows $\mu/E_0$ dependence of $T_{\rm c}$ at $m_2/m_1=1.3$.  
    $\tilde{U}_{12}=10^{-3}$ and $\Lambda/k_0=10$ are used. For comparison, the dotted curves show the BCS results without the GMB correction.
    The lower panels represent the ratio between the superconducting critical temperatures with and without the GMB corrections as functions of (b1) $m_2/m_1$ at $\mu/E_0=0.6$ and (b2) $\mu/E_0$ at $m_2/m_1=1.3$.
    The horizontal dashed line indicates the ratio $(4e)^{-1/3}\simeq 0.45$ in the single-band counterpart at weak coupling.
    \red{The value at $m_2/m_1 = 1$ is marked with the horizontal thin chain-dotted line in (b1).}
    }
    \label{fig:3}
\end{figure}

A notable feature in Fig,~\ref{fig:2} is that $T_{\rm c}$ is much larger for $m_2/m_1\gesim 10$ than for $m_2/m_1\simeq 1$ even with the strong GMB reduction.  Let us examine this more closely in 
Fig.~\ref{fig:3}(a1), which compares the $m_2/m_1$-dependence of $T_{\rm c}$ between BCS and GMB at $\mu/E_0=0.6$ with $\tilde{U}_{12}=10^{-3}$. While the BCS result has the saturation of $T_{\rm c}$ at larger $m_2/m_1$ as expected from Eq.~\eqref{eq:gapeq_largemass},
the GMB result exhibits a {\it peak} of $T_{\rm c}$ around a finite $m_2/m_1=3.5$, beyond which $T_{\rm c}$ decreases monotonically with $m_2/m_1$.
We can interpret the remarkable result as signifying a {\it competition} between the enhanced pairing due to the strong attraction in Eq.~\eqref{eq:gapeq_largemass} and the strong GMB reduction. 
The tradeoff results in an optimal mass ratio, which depends on the momentum cutoff ($\sim$ band width in lattice models) in the incipient heavy band, but 
the peaked structure persists when $\Lambda$ is varied; see  Supplement~\cite{Supplement}.
To accurately evaluate the 
cutoff dependence, we would have to adopt some kind of renormalization 
scheme, which will be an interesting future work.
We note that $T_{\rm c }$ does not exhibit a peak in the $\mu/E_0$-dependence as shown in Fig.~\ref{fig:3}(a2).

\noindent
\textit{Suppressed particle-hole fluctuations near the Fano-Feshbach resonance}---
We can further capture the behavior of $T_{\rm c}$ in 
terms of an underlying resonance.  For that, let us look at 
the ratio $T_{\rm c}^{\rm GMB}/T_{\rm c}^{\rm BCS}$ between BCS and GMB 
schemes in Figs.~\ref{fig:3}(b1) and (b2).
\red{This ratio measures the extent to which the GMB screening is at work.}
In both of $m_2/m_1$ and $\mu/E_0$ dependencies,
$T_{\rm c}^{\rm GMB}/T_{\rm c}^{\rm BCS}$ exhibits a peaked behavior.
Around $m_2/m_1=1.0$, one can find $T_{\rm c}^{\rm GMB}/T_{\rm c}^{\rm BCS}\simeq 0.4$ (for $k_0a_{11}=-1.0$ here) 
regardless of the value of $\tilde{U}_{12}$, which originates from the GMB screening associated with the Fermi surface in band $1$.  
Similar values $T_{\rm c}^{\rm GMB}/T_{\rm c}^{\rm BCS}\simeq 0.4 - 0.5$ are reported for $k_{\rm F}|a|<1.0$ in a single-band study~\cite{PhysRevB.97.014528}.  

As $m_2/m_1$ is increased, we have a conspicuous peak \red{of $T_{\rm c}^{\rm GMB}/T_{\rm c}^{\rm BCS}$}, after which 
$T_{\rm c}^{\rm GMB}$ starts to decrease because of enhanced particle-hole fluctuations for larger $m_2/m_1$, 
and the ratio eventually drops below the single-band GMB result at weak coupling given by $(4e)^{-1/3}\simeq 0.45$. 
Nevertheless, the GMB reduction for larger $m_2/m_1$  is not drastic, \red{indicating that the enhanced pairing effect is still remarkable at $m_2/m_1\gesim 10$. Thus, 
even in the presence of the GMB correction, $T_{\rm c}$ remains large at $m_2/m_1\gesim 10$ as compared to the case for $m_2/m_1\simeq 1$ in Fig~\ref{fig:3}(a1).}

\begin{figure}[t]
    \centering
    \includegraphics[width=7cm]{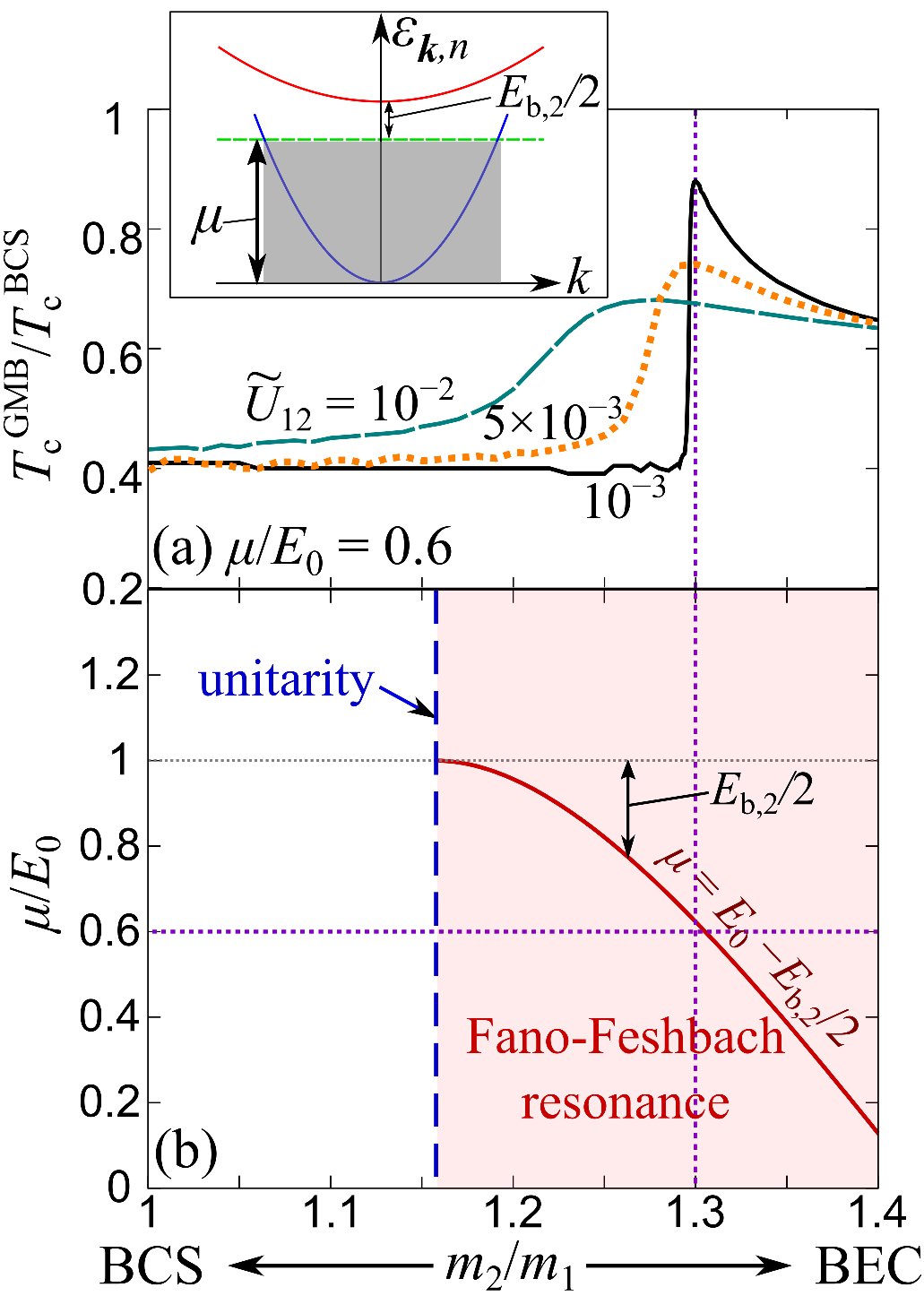}
    \caption{
    (a) Calculated $T_{\rm c}^{\rm GMB}/T_{\rm c}^{\rm BCS}$ as a function of $m_2/m_1$ with $\tilde{U}_{12}=10^{-2}$, $5\times 10^{-3}$, and $10^{-3}$ at $\mu/E_0=0.6$.
    The inset shows the schematics for single-particle energy level with the chemical potential $\mu$ touching the Fano-Feshbach resonance associated with the incipient heavy band.
    (b) Fano-Feshbach resonance line $\mu=E_0-E_{\rm b,2}/2$ at the weak interband coupling limit ($U_{12}\rightarrow 0$). The vertical blue line indicates the unitarity at which the two-body bound state appears in band 2.  
    \red{One can assume that the BCS (BEC) regime is realized in band 2 when $m_2/m_1$ is small (large).}
    Purple lines mark the case of $\mu/E_0=0.6$ considered in (a).
    }
    \label{fig:4}
\end{figure}

Now, let us analyze the peak in 
$T_{\rm c}^{\rm GMB}/T_{\rm c}^{\rm BCS}$ against $m_2/m_1$ in 
Fig.~\ref{fig:4}(a) for various values of $\tilde{U}_{12}=10^{-3}$, $5\times 10^{-3}$, and $10^{-2}$ at $\mu/E_0=0.6$.
When $m_2/m_1$ increased from $1$, a sharp enhancement of $T_{\rm c}^{\rm GMB}/T_{\rm c}^{\rm BCS}$ emerges, especially at $\tilde{U}_{12}=10^{-3}$ in Fig.~\ref{fig:4}(a), which indicates that the GMB correction on $T_{\rm c}$ is dramatically reduced there.  
The peak starts to be smeared for larger  $\tilde{U}_{12}$.

Another notable feature is that the peak has 
an asymmetric shape in Fig.~\ref{fig:4}(a), 
which we can immediately recognize as reminiscent of the Fano-Feshbach resonance.
Indeed, physics behind the dramatic reduction of the GMB correction on $T_{\rm c}$ for $\mu/E_0<1.0$ and smaller $\tilde{U}_{12}$ revealed in Fig.~\ref{fig:4}(a) should be a consequence of the chemical potential touching the incipient heavy band, 
thereby causing a Fano-Feshbach resonance in the following sense. 
The heavy band accommodates a bound state (which turns into the resonance state for nonzero $\tilde{U}_{12}$) for $m_2/m_1\geq 1-\frac{\pi}{2a_{11}\Lambda}\; (\simeq 1.16$ for the present choice of $k_0 a_{11}=-1.0$ and $\Lambda/k_0=10$)~\cite{Supplement}.
\red{$m_2/m_1\simeq 1.16$ can be regarded as the unitarity,~\cite{strinati2018bcs,OHASHI2020103739}
and the geometrical control of $m_2/m_1$, e.g., by band engineering with quantum confinement or orbital selection, leads to the BCS-BEC crossover as indicated in Fig.~\ref{fig:4}(b).}
For small $U_{12}$, the resonance energy $\omega_{\rm res}$ is given by 
\begin{align}
    \omega_{\rm res}=-E_{\rm b,2}+2E_0+O(U_{12}^4),
\end{align}
where $E_{\rm b,2}$ is the two-body binding energy 
(inset of Fig.~\ref{fig:4}(a)) in band $2$ for $U_{12}\rightarrow 0$. 
We can see that the peak of $T_{\rm c}^{\rm GMB}/T_{\rm c}^{\rm BCS}$ for $\tilde{U}_{12}=10^{-3}$ and $\mu/E_0<1.0$ does indeed take place at the mass ratio at which the Fano-Feshbach resonance resides, whose position shifts as $\mu/E_0$ is varied.  
Namely, the resonance arises at $\mu=\omega_{\rm res}/2\simeq E_0-E_{\rm b,2}/2$ (with $1/2$ 
for putting the two-body energy into the one per particle).

We have actually plotted in Fig.~\ref{fig:4}(b) the trajectory $\mu=E_0-E_{\rm b,2}/2$ against $m_2/m_1$ as 
the Fano-Feshbach resonance line.   
We can see that the sharp reduction of the GMB correction occurs right at the resonance.  
Both of the bound state and the resonance 
start to exist above the unitarity mass ratio $m_2/m_1=1-\frac{\pi}{2a_{11}\Lambda}$~\cite{Supplement} where the two-body bound state appears in band 2.
When $E_{\rm b,2}$ arises, as depicted schematically 
in Fig.~\ref{fig:4}(a) and marked with a double arrow in (b), 
electrons primarily occupy the resonant state, while the second band is basically empty for $\mu-E_0<0$ (except for thermally excited quasiparticles).
In such a case, the heavy band is in an extremely dilute (i.e., strong-coupling) regime characterized by $(\mu-E_0)/E_{\rm b,2}\simeq -1/2$,  which 
is a counterpart to the single-band expression,  $\mu/E_{\rm b}=-1/2$, for the chemical potential in the BEC limit (with $E_{\rm b}$ being the  binding energy in the single-band case).   
The realization of the strong-coupling limit and the verge 
of appearance of the second-band Fermi surface, taking place 
 around the Fano-Feshbach resonance at $\mu=E_0-E_{\rm b,2}/2$, 
 thus lead to the suppression of the GMB screening effect.

\noindent
\textit{Summary}---
We have investigated the GMB screening effect on the superconducting critical temperature in a two-band superconductor consisting of a deep dispersive (light-mass) band and a heavy-mass band with the chemical potential adjusted to make heavy band incipient.
By developing the diagrammatic GMB formalism for two-band systems,
we have calculated the superconducting critical temperature $T_{\rm c}$ for various values of (i) the mass ratio of the two bands, (ii) chemical potential, and (iii) the pair-exchange coupling.
A strong reduction of $T_{\rm c}$, which we 
traced back to extremely large particle-hole fluctuations when the second band has a heavy mass, is found to be overcome, 
because the GMB reduction 
has to {\it compete} with the enhanced pairing interaction arising from the incipient heavy band, resulting in a {\it peaked} structure in $T_{\rm c}$ versus the mass ratio.  We have then unraveled that there indeed exists 
a Fano-Feshbach resonance that occurs  
when the chemical potential traverses 
the energy of the two-body bound state emerging 
below the heavy 
band (which becomes a resonant state in the presence of the interband pair exchange). The GMB diagram is strongly
suppressed when the Fermi surface of the incipient heavy band is collapsed because of the bound state leading to the Fano-Feshbach resonance. Thus we end up with a mechanism for evading the screening effects of particle-hole (GMB) fluctuations, leaving the critical temperature in a protectorate regime of parameters.

The present results are expected to give a hint for further understanding of many-body physics in multi-component condensations as well as material design toward high-$T_{\rm c}$ superconductors with band
or structure engineering such as superlattices.
As a future perspective,
it would be interesting to go beyond the present approach by incorporating the full momentum- and energy-dependence of the particle-hole diagrams and the Popov correction for interacting molecular pairs, following the approach of Ref.~\cite{PhysRevB.97.014528}.
The effects of spin-orbit coupling may also be important in applying the present approach to 
topological superconductors with Rashba heterostructures~\cite{PhysRevB.103.024523}.
We can mention in passing that, in lattice systems where the particle-hole transformation can be applied in certain conditions,
it could be possible, through the attraction-repulsion transformation, to clarify the relevance of the present scheme to the repulsive multi-band systems where spin fluctuations are dominant~\cite{PhysRevB.42.2125,PhysRevLett.69.3820,PhysRevB.48.7598,note,PhysRevB.101.014501,Sayyad_2023,PhysRevResearch.2.033356}. 
Lattices also make the 
introduction of cutoffs unnecessary, which will facilitate the diagrammatic analysis.
\red{It is also worth studying the role of the low-dimensionality such as the GMB effect on the Berezinskii-Kosterlitz-Thouless transition~\cite{kosterlitz2016kosterlitz,midei2024predictive,paramasivam2023high} and on the behavior of the suppression coefficient of the mean-field pairing temperature for two-dimensional systems, which are of considerable interest.}

\noindent\textit{Acknowledgements}---
H.~T. thanks Y.~Yerin, P.~Pieri, K.~Ochi, K.~Iida, and H.~Liang group in Univ. Tokyo for the useful discussion.
H.~T.
was supported by the JSPS Grants-in-Aid for Scientific Research under Grants No.~18H05406, No.~22H01158, and No.~22K13981.  
H.~A. thanks CREST (Grant Number JPMJCR18T4).
P.~A. was supported by PNRR MUR project PE0000023-NQSTI.

\bibliographystyle{apsrev4-2}
\bibliography{reference.bib}

\begin{thebibliography}{65}%
\makeatletter
\providecommand \@ifxundefined [1]{%
 \@ifx{#1\undefined}
}%
\providecommand \@ifnum [1]{%
 \ifnum #1\expandafter \@firstoftwo
 \else \expandafter \@secondoftwo
 \fi
}%
\providecommand \@ifx [1]{%
 \ifx #1\expandafter \@firstoftwo
 \else \expandafter \@secondoftwo
 \fi
}%
\providecommand \natexlab [1]{#1}%
\providecommand \enquote  [1]{``#1''}%
\providecommand \bibnamefont  [1]{#1}%
\providecommand \bibfnamefont [1]{#1}%
\providecommand \citenamefont [1]{#1}%
\providecommand \href@noop [0]{\@secondoftwo}%
\providecommand \href [0]{\begingroup \@sanitize@url \@href}%
\providecommand \@href[1]{\@@startlink{#1}\@@href}%
\providecommand \@@href[1]{\endgroup#1\@@endlink}%
\providecommand \@sanitize@url [0]{\catcode `\\12\catcode `\$12\catcode `\&12\catcode `\#12\catcode `\^12\catcode `\_12\catcode `\%12\relax}%
\providecommand \@@startlink[1]{}%
\providecommand \@@endlink[0]{}%
\providecommand \url  [0]{\begingroup\@sanitize@url \@url }%
\providecommand \@url [1]{\endgroup\@href {#1}{\urlprefix }}%
\providecommand \urlprefix  [0]{URL }%
\providecommand \Eprint [0]{\href }%
\providecommand \doibase [0]{https://doi.org/}%
\providecommand \selectlanguage [0]{\@gobble}%
\providecommand \bibinfo  [0]{\@secondoftwo}%
\providecommand \bibfield  [0]{\@secondoftwo}%
\providecommand \translation [1]{[#1]}%
\providecommand \BibitemOpen [0]{}%
\providecommand \bibitemStop [0]{}%
\providecommand \bibitemNoStop [0]{.\EOS\space}%
\providecommand \EOS [0]{\spacefactor3000\relax}%
\providecommand \BibitemShut  [1]{\csname bibitem#1\endcsname}%
\let\auto@bib@innerbib\@empty
\bibitem [{\citenamefont {Bohr}\ \emph {et~al.}(1958)\citenamefont {Bohr}, \citenamefont {Mottelson},\ and\ \citenamefont {Pines}}]{PhysRev.110.936}%
  \BibitemOpen
  \bibfield  {author} {\bibinfo {author} {\bibfnamefont {A.}~\bibnamefont {Bohr}}, \bibinfo {author} {\bibfnamefont {B.~R.}\ \bibnamefont {Mottelson}},\ and\ \bibinfo {author} {\bibfnamefont {D.}~\bibnamefont {Pines}},\ }\href {https://doi.org/10.1103/PhysRev.110.936} {\bibfield  {journal} {\bibinfo  {journal} {Phys. Rev.}\ }\textbf {\bibinfo {volume} {110}},\ \bibinfo {pages} {936} (\bibinfo {year} {1958})}\BibitemShut {NoStop}%
\bibitem [{\citenamefont {Nambu}\ and\ \citenamefont {Jona-Lasinio}(1961{\natexlab{a}})}]{PhysRev.122.345}%
  \BibitemOpen
  \bibfield  {author} {\bibinfo {author} {\bibfnamefont {Y.}~\bibnamefont {Nambu}}\ and\ \bibinfo {author} {\bibfnamefont {G.}~\bibnamefont {Jona-Lasinio}},\ }\href {https://doi.org/10.1103/PhysRev.122.345} {\bibfield  {journal} {\bibinfo  {journal} {Phys. Rev.}\ }\textbf {\bibinfo {volume} {122}},\ \bibinfo {pages} {345} (\bibinfo {year} {1961}{\natexlab{a}})}\BibitemShut {NoStop}%
\bibitem [{\citenamefont {Nambu}\ and\ \citenamefont {Jona-Lasinio}(1961{\natexlab{b}})}]{PhysRev.124.246}%
  \BibitemOpen
  \bibfield  {author} {\bibinfo {author} {\bibfnamefont {Y.}~\bibnamefont {Nambu}}\ and\ \bibinfo {author} {\bibfnamefont {G.}~\bibnamefont {Jona-Lasinio}},\ }\href {https://doi.org/10.1103/PhysRev.124.246} {\bibfield  {journal} {\bibinfo  {journal} {Phys. Rev.}\ }\textbf {\bibinfo {volume} {124}},\ \bibinfo {pages} {246} (\bibinfo {year} {1961}{\natexlab{b}})}\BibitemShut {NoStop}%
\bibitem [{\citenamefont {Bednorz}\ and\ \citenamefont {M{\"u}ller}(1986)}]{bednorz1986possible}%
  \BibitemOpen
  \bibfield  {author} {\bibinfo {author} {\bibfnamefont {J.~G.}\ \bibnamefont {Bednorz}}\ and\ \bibinfo {author} {\bibfnamefont {K.~A.}\ \bibnamefont {M{\"u}ller}},\ }\href@noop {} {\bibfield  {journal} {\bibinfo  {journal} {Zeitschrift f{\"u}r Physik B Condensed Matter}\ }\textbf {\bibinfo {volume} {64}},\ \bibinfo {pages} {189} (\bibinfo {year} {1986})}\BibitemShut {NoStop}%
\bibitem [{\citenamefont {Kamihara}\ \emph {et~al.}(2008)\citenamefont {Kamihara}, \citenamefont {Watanabe}, \citenamefont {Hirano},\ and\ \citenamefont {Hosono}}]{kamihara2008iron}%
  \BibitemOpen
  \bibfield  {author} {\bibinfo {author} {\bibfnamefont {Y.}~\bibnamefont {Kamihara}}, \bibinfo {author} {\bibfnamefont {T.}~\bibnamefont {Watanabe}}, \bibinfo {author} {\bibfnamefont {M.}~\bibnamefont {Hirano}},\ and\ \bibinfo {author} {\bibfnamefont {H.}~\bibnamefont {Hosono}},\ }\href@noop {} {\bibfield  {journal} {\bibinfo  {journal} {J. American Chem. Soc.}\ }\textbf {\bibinfo {volume} {130}},\ \bibinfo {pages} {3296} (\bibinfo {year} {2008})}\BibitemShut {NoStop}%
\bibitem [{\citenamefont {Eagles}(1969)}]{PhysRev.186.456}%
  \BibitemOpen
  \bibfield  {author} {\bibinfo {author} {\bibfnamefont {D.~M.}\ \bibnamefont {Eagles}},\ }\href {https://doi.org/10.1103/PhysRev.186.456} {\bibfield  {journal} {\bibinfo  {journal} {Phys. Rev.}\ }\textbf {\bibinfo {volume} {186}},\ \bibinfo {pages} {456} (\bibinfo {year} {1969})}\BibitemShut {NoStop}%
\bibitem [{\citenamefont {Leggett}(2008)}]{leggett2008diatomic}%
  \BibitemOpen
  \bibfield  {author} {\bibinfo {author} {\bibfnamefont {A.~J.}\ \bibnamefont {Leggett}},\ }in\ \href@noop {} {\emph {\bibinfo {booktitle} {Modern Trends in the Theory of Condensed Matter: Proc. XVI Karpacz Winter School of Theoretical Physics, 1979, Karpacz, Poland}}}\ (\bibinfo  {publisher} {Springer},\ \bibinfo {year} {2008})\ pp.\ \bibinfo {pages} {13--27}\BibitemShut {NoStop}%
\bibitem [{\citenamefont {Nozieres}\ and\ \citenamefont {Schmitt-Rink}(1985)}]{nozieres1985bose}%
  \BibitemOpen
  \bibfield  {author} {\bibinfo {author} {\bibfnamefont {P.}~\bibnamefont {Nozieres}}\ and\ \bibinfo {author} {\bibfnamefont {S.}~\bibnamefont {Schmitt-Rink}},\ }\href@noop {} {\bibfield  {journal} {\bibinfo  {journal} {Journal of Low Temperature Physics}\ }\textbf {\bibinfo {volume} {59}},\ \bibinfo {pages} {195} (\bibinfo {year} {1985})}\BibitemShut {NoStop}%
\bibitem [{\citenamefont {S\'a~de Melo}\ \emph {et~al.}(1993)\citenamefont {S\'a~de Melo}, \citenamefont {Randeria},\ and\ \citenamefont {Engelbrecht}}]{PhysRevLett.71.3202}%
  \BibitemOpen
  \bibfield  {author} {\bibinfo {author} {\bibfnamefont {C.~A.~R.}\ \bibnamefont {S\'a~de Melo}}, \bibinfo {author} {\bibfnamefont {M.}~\bibnamefont {Randeria}},\ and\ \bibinfo {author} {\bibfnamefont {J.~R.}\ \bibnamefont {Engelbrecht}},\ }\href {https://doi.org/10.1103/PhysRevLett.71.3202} {\bibfield  {journal} {\bibinfo  {journal} {Phys. Rev. Lett.}\ }\textbf {\bibinfo {volume} {71}},\ \bibinfo {pages} {3202} (\bibinfo {year} {1993})}\BibitemShut {NoStop}%
\bibitem [{\citenamefont {Chin}\ \emph {et~al.}(2010)\citenamefont {Chin}, \citenamefont {Grimm}, \citenamefont {Julienne},\ and\ \citenamefont {Tiesinga}}]{RevModPhys.82.1225}%
  \BibitemOpen
  \bibfield  {author} {\bibinfo {author} {\bibfnamefont {C.}~\bibnamefont {Chin}}, \bibinfo {author} {\bibfnamefont {R.}~\bibnamefont {Grimm}}, \bibinfo {author} {\bibfnamefont {P.}~\bibnamefont {Julienne}},\ and\ \bibinfo {author} {\bibfnamefont {E.}~\bibnamefont {Tiesinga}},\ }\href {https://doi.org/10.1103/RevModPhys.82.1225} {\bibfield  {journal} {\bibinfo  {journal} {Rev. Mod. Phys.}\ }\textbf {\bibinfo {volume} {82}},\ \bibinfo {pages} {1225} (\bibinfo {year} {2010})}\BibitemShut {NoStop}%
\bibitem [{\citenamefont {Regal}\ \emph {et~al.}(2004)\citenamefont {Regal}, \citenamefont {Greiner},\ and\ \citenamefont {Jin}}]{PhysRevLett.92.040403}%
  \BibitemOpen
  \bibfield  {author} {\bibinfo {author} {\bibfnamefont {C.~A.}\ \bibnamefont {Regal}}, \bibinfo {author} {\bibfnamefont {M.}~\bibnamefont {Greiner}},\ and\ \bibinfo {author} {\bibfnamefont {D.~S.}\ \bibnamefont {Jin}},\ }\href {https://doi.org/10.1103/PhysRevLett.92.040403} {\bibfield  {journal} {\bibinfo  {journal} {Phys. Rev. Lett.}\ }\textbf {\bibinfo {volume} {92}},\ \bibinfo {pages} {040403} (\bibinfo {year} {2004})}\BibitemShut {NoStop}%
\bibitem [{\citenamefont {Zwierlein}\ \emph {et~al.}(2004)\citenamefont {Zwierlein}, \citenamefont {Stan}, \citenamefont {Schunck}, \citenamefont {Raupach}, \citenamefont {Kerman},\ and\ \citenamefont {Ketterle}}]{PhysRevLett.92.120403}%
  \BibitemOpen
  \bibfield  {author} {\bibinfo {author} {\bibfnamefont {M.~W.}\ \bibnamefont {Zwierlein}}, \bibinfo {author} {\bibfnamefont {C.~A.}\ \bibnamefont {Stan}}, \bibinfo {author} {\bibfnamefont {C.~H.}\ \bibnamefont {Schunck}}, \bibinfo {author} {\bibfnamefont {S.~M.~F.}\ \bibnamefont {Raupach}}, \bibinfo {author} {\bibfnamefont {A.~J.}\ \bibnamefont {Kerman}},\ and\ \bibinfo {author} {\bibfnamefont {W.}~\bibnamefont {Ketterle}},\ }\href {https://doi.org/10.1103/PhysRevLett.92.120403} {\bibfield  {journal} {\bibinfo  {journal} {Phys. Rev. Lett.}\ }\textbf {\bibinfo {volume} {92}},\ \bibinfo {pages} {120403} (\bibinfo {year} {2004})}\BibitemShut {NoStop}%
\bibitem [{\citenamefont {Bartenstein}\ \emph {et~al.}(2004)\citenamefont {Bartenstein}, \citenamefont {Altmeyer}, \citenamefont {Riedl}, \citenamefont {Jochim}, \citenamefont {Chin}, \citenamefont {Denschlag},\ and\ \citenamefont {Grimm}}]{PhysRevLett.92.203201}%
  \BibitemOpen
  \bibfield  {author} {\bibinfo {author} {\bibfnamefont {M.}~\bibnamefont {Bartenstein}}, \bibinfo {author} {\bibfnamefont {A.}~\bibnamefont {Altmeyer}}, \bibinfo {author} {\bibfnamefont {S.}~\bibnamefont {Riedl}}, \bibinfo {author} {\bibfnamefont {S.}~\bibnamefont {Jochim}}, \bibinfo {author} {\bibfnamefont {C.}~\bibnamefont {Chin}}, \bibinfo {author} {\bibfnamefont {J.~H.}\ \bibnamefont {Denschlag}},\ and\ \bibinfo {author} {\bibfnamefont {R.}~\bibnamefont {Grimm}},\ }\href {https://doi.org/10.1103/PhysRevLett.92.203201} {\bibfield  {journal} {\bibinfo  {journal} {Phys. Rev. Lett.}\ }\textbf {\bibinfo {volume} {92}},\ \bibinfo {pages} {203201} (\bibinfo {year} {2004})}\BibitemShut {NoStop}%
\bibitem [{\citenamefont {Lubashevsky}\ \emph {et~al.}(2012)\citenamefont {Lubashevsky}, \citenamefont {Lahoud}, \citenamefont {Chashka}, \citenamefont {Podolsky},\ and\ \citenamefont {Kanigel}}]{lubashevsky2012shallow}%
  \BibitemOpen
  \bibfield  {author} {\bibinfo {author} {\bibfnamefont {Y.}~\bibnamefont {Lubashevsky}}, \bibinfo {author} {\bibfnamefont {E.}~\bibnamefont {Lahoud}}, \bibinfo {author} {\bibfnamefont {K.}~\bibnamefont {Chashka}}, \bibinfo {author} {\bibfnamefont {D.}~\bibnamefont {Podolsky}},\ and\ \bibinfo {author} {\bibfnamefont {A.}~\bibnamefont {Kanigel}},\ }\href@noop {} {\bibfield  {journal} {\bibinfo  {journal} {Nature Physics}\ }\textbf {\bibinfo {volume} {8}},\ \bibinfo {pages} {309} (\bibinfo {year} {2012})}\BibitemShut {NoStop}%
\bibitem [{\citenamefont {Kasahara}\ \emph {et~al.}(2014)\citenamefont {Kasahara}, \citenamefont {Watashige}, \citenamefont {Hanaguri}, \citenamefont {Kohsaka}, \citenamefont {Yamashita}, \citenamefont {Shimoyama}, \citenamefont {Mizukami}, \citenamefont {Endo}, \citenamefont {Ikeda}, \citenamefont {Aoyama} \emph {et~al.}}]{kasahara2014field}%
  \BibitemOpen
  \bibfield  {author} {\bibinfo {author} {\bibfnamefont {S.}~\bibnamefont {Kasahara}}, \bibinfo {author} {\bibfnamefont {T.}~\bibnamefont {Watashige}}, \bibinfo {author} {\bibfnamefont {T.}~\bibnamefont {Hanaguri}}, \bibinfo {author} {\bibfnamefont {Y.}~\bibnamefont {Kohsaka}}, \bibinfo {author} {\bibfnamefont {T.}~\bibnamefont {Yamashita}}, \bibinfo {author} {\bibfnamefont {Y.}~\bibnamefont {Shimoyama}}, \bibinfo {author} {\bibfnamefont {Y.}~\bibnamefont {Mizukami}}, \bibinfo {author} {\bibfnamefont {R.}~\bibnamefont {Endo}}, \bibinfo {author} {\bibfnamefont {H.}~\bibnamefont {Ikeda}}, \bibinfo {author} {\bibfnamefont {K.}~\bibnamefont {Aoyama}}, \emph {et~al.},\ }\href@noop {} {\bibfield  {journal} {\bibinfo  {journal} {Proceedings of the National Academy of Sciences}\ }\textbf {\bibinfo {volume} {111}},\ \bibinfo {pages} {16309} (\bibinfo {year} {2014})}\BibitemShut {NoStop}%
\bibitem [{\citenamefont {Rinott}\ \emph {et~al.}(2017)\citenamefont {Rinott}, \citenamefont {Chashka}, \citenamefont {Ribak}, \citenamefont {Rienks}, \citenamefont {Taleb-Ibrahimi}, \citenamefont {Fevre}, \citenamefont {Bertran}, \citenamefont {Randeria},\ and\ \citenamefont {Kanigel}}]{doi:10.1126/sciadv.1602372}%
  \BibitemOpen
  \bibfield  {author} {\bibinfo {author} {\bibfnamefont {S.}~\bibnamefont {Rinott}}, \bibinfo {author} {\bibfnamefont {K.~B.}\ \bibnamefont {Chashka}}, \bibinfo {author} {\bibfnamefont {A.}~\bibnamefont {Ribak}}, \bibinfo {author} {\bibfnamefont {E.~D.~L.}\ \bibnamefont {Rienks}}, \bibinfo {author} {\bibfnamefont {A.}~\bibnamefont {Taleb-Ibrahimi}}, \bibinfo {author} {\bibfnamefont {P.~L.}\ \bibnamefont {Fevre}}, \bibinfo {author} {\bibfnamefont {F.}~\bibnamefont {Bertran}}, \bibinfo {author} {\bibfnamefont {M.}~\bibnamefont {Randeria}},\ and\ \bibinfo {author} {\bibfnamefont {A.}~\bibnamefont {Kanigel}},\ }\href {https://doi.org/10.1126/sciadv.1602372} {\bibfield  {journal} {\bibinfo  {journal} {Science Advances}\ }\textbf {\bibinfo {volume} {3}},\ \bibinfo {pages} {e1602372} (\bibinfo {year} {2017})}\BibitemShut {NoStop}%
\bibitem [{\citenamefont {Hashimoto}\ \emph {et~al.}(2020)\citenamefont {Hashimoto}, \citenamefont {Ota}, \citenamefont {Tsuzuki}, \citenamefont {Nagashima}, \citenamefont {Fukushima}, \citenamefont {Kasahara}, \citenamefont {Matsuda}, \citenamefont {Matsuura}, \citenamefont {Mizukami}, \citenamefont {Shibauchi}, \citenamefont {Shin},\ and\ \citenamefont {Okazaki}}]{doi:10.1126/sciadv.abb9052}%
  \BibitemOpen
  \bibfield  {author} {\bibinfo {author} {\bibfnamefont {T.}~\bibnamefont {Hashimoto}}, \bibinfo {author} {\bibfnamefont {Y.}~\bibnamefont {Ota}}, \bibinfo {author} {\bibfnamefont {A.}~\bibnamefont {Tsuzuki}}, \bibinfo {author} {\bibfnamefont {T.}~\bibnamefont {Nagashima}}, \bibinfo {author} {\bibfnamefont {A.}~\bibnamefont {Fukushima}}, \bibinfo {author} {\bibfnamefont {S.}~\bibnamefont {Kasahara}}, \bibinfo {author} {\bibfnamefont {Y.}~\bibnamefont {Matsuda}}, \bibinfo {author} {\bibfnamefont {K.}~\bibnamefont {Matsuura}}, \bibinfo {author} {\bibfnamefont {Y.}~\bibnamefont {Mizukami}}, \bibinfo {author} {\bibfnamefont {T.}~\bibnamefont {Shibauchi}}, \bibinfo {author} {\bibfnamefont {S.}~\bibnamefont {Shin}},\ and\ \bibinfo {author} {\bibfnamefont {K.}~\bibnamefont {Okazaki}},\ }\href {https://doi.org/10.1126/sciadv.abb9052} {\bibfield  {journal} {\bibinfo  {journal} {Science Advances}\ }\textbf {\bibinfo {volume} {6}},\ \bibinfo {pages} {eabb9052} (\bibinfo {year} {2020})}\BibitemShut {NoStop}%
\bibitem [{\citenamefont {Mizukami}\ \emph {et~al.}(2023)\citenamefont {Mizukami}, \citenamefont {Haze}, \citenamefont {Tanaka}, \citenamefont {Matsuura}, \citenamefont {Sano}, \citenamefont {B{\"o}ker}, \citenamefont {Eremin}, \citenamefont {Kasahara}, \citenamefont {Matsuda},\ and\ \citenamefont {Shibauchi}}]{mizukami2023unusual}%
  \BibitemOpen
  \bibfield  {author} {\bibinfo {author} {\bibfnamefont {Y.}~\bibnamefont {Mizukami}}, \bibinfo {author} {\bibfnamefont {M.}~\bibnamefont {Haze}}, \bibinfo {author} {\bibfnamefont {O.}~\bibnamefont {Tanaka}}, \bibinfo {author} {\bibfnamefont {K.}~\bibnamefont {Matsuura}}, \bibinfo {author} {\bibfnamefont {D.}~\bibnamefont {Sano}}, \bibinfo {author} {\bibfnamefont {J.}~\bibnamefont {B{\"o}ker}}, \bibinfo {author} {\bibfnamefont {I.}~\bibnamefont {Eremin}}, \bibinfo {author} {\bibfnamefont {S.}~\bibnamefont {Kasahara}}, \bibinfo {author} {\bibfnamefont {Y.}~\bibnamefont {Matsuda}},\ and\ \bibinfo {author} {\bibfnamefont {T.}~\bibnamefont {Shibauchi}},\ }\href@noop {} {\bibfield  {journal} {\bibinfo  {journal} {Commun. Phys.}\ }\textbf {\bibinfo {volume} {6}},\ \bibinfo {pages} {183} (\bibinfo {year} {2023})}\BibitemShut {NoStop}%
\bibitem [{\citenamefont {Nakagawa}\ \emph {et~al.}(2018)\citenamefont {Nakagawa}, \citenamefont {Saito}, \citenamefont {Nojima}, \citenamefont {Inumaru}, \citenamefont {Yamanaka}, \citenamefont {Kasahara},\ and\ \citenamefont {Iwasa}}]{PhysRevB.98.064512}%
  \BibitemOpen
  \bibfield  {author} {\bibinfo {author} {\bibfnamefont {Y.}~\bibnamefont {Nakagawa}}, \bibinfo {author} {\bibfnamefont {Y.}~\bibnamefont {Saito}}, \bibinfo {author} {\bibfnamefont {T.}~\bibnamefont {Nojima}}, \bibinfo {author} {\bibfnamefont {K.}~\bibnamefont {Inumaru}}, \bibinfo {author} {\bibfnamefont {S.}~\bibnamefont {Yamanaka}}, \bibinfo {author} {\bibfnamefont {Y.}~\bibnamefont {Kasahara}},\ and\ \bibinfo {author} {\bibfnamefont {Y.}~\bibnamefont {Iwasa}},\ }\href {https://doi.org/10.1103/PhysRevB.98.064512} {\bibfield  {journal} {\bibinfo  {journal} {Phys. Rev. B}\ }\textbf {\bibinfo {volume} {98}},\ \bibinfo {pages} {064512} (\bibinfo {year} {2018})}\BibitemShut {NoStop}%
\bibitem [{\citenamefont {Nakagawa}\ \emph {et~al.}(2021)\citenamefont {Nakagawa}, \citenamefont {Kasahara}, \citenamefont {Nomoto}, \citenamefont {Arita}, \citenamefont {Nojima},\ and\ \citenamefont {Iwasa}}]{nakagawa2021gate}%
  \BibitemOpen
  \bibfield  {author} {\bibinfo {author} {\bibfnamefont {Y.}~\bibnamefont {Nakagawa}}, \bibinfo {author} {\bibfnamefont {Y.}~\bibnamefont {Kasahara}}, \bibinfo {author} {\bibfnamefont {T.}~\bibnamefont {Nomoto}}, \bibinfo {author} {\bibfnamefont {R.}~\bibnamefont {Arita}}, \bibinfo {author} {\bibfnamefont {T.}~\bibnamefont {Nojima}},\ and\ \bibinfo {author} {\bibfnamefont {Y.}~\bibnamefont {Iwasa}},\ }\href@noop {} {\bibfield  {journal} {\bibinfo  {journal} {Science}\ }\textbf {\bibinfo {volume} {372}},\ \bibinfo {pages} {190} (\bibinfo {year} {2021})}\BibitemShut {NoStop}%
\bibitem [{\citenamefont {Suzuki}\ \emph {et~al.}(2022)\citenamefont {Suzuki}, \citenamefont {Wakamatsu}, \citenamefont {Ibuka}, \citenamefont {Oike}, \citenamefont {Fujii}, \citenamefont {Miyagawa}, \citenamefont {Taniguchi},\ and\ \citenamefont {Kanoda}}]{PhysRevX.12.011016}%
  \BibitemOpen
  \bibfield  {author} {\bibinfo {author} {\bibfnamefont {Y.}~\bibnamefont {Suzuki}}, \bibinfo {author} {\bibfnamefont {K.}~\bibnamefont {Wakamatsu}}, \bibinfo {author} {\bibfnamefont {J.}~\bibnamefont {Ibuka}}, \bibinfo {author} {\bibfnamefont {H.}~\bibnamefont {Oike}}, \bibinfo {author} {\bibfnamefont {T.}~\bibnamefont {Fujii}}, \bibinfo {author} {\bibfnamefont {K.}~\bibnamefont {Miyagawa}}, \bibinfo {author} {\bibfnamefont {H.}~\bibnamefont {Taniguchi}},\ and\ \bibinfo {author} {\bibfnamefont {K.}~\bibnamefont {Kanoda}},\ }\href {https://doi.org/10.1103/PhysRevX.12.011016} {\bibfield  {journal} {\bibinfo  {journal} {Phys. Rev. X}\ }\textbf {\bibinfo {volume} {12}},\ \bibinfo {pages} {011016} (\bibinfo {year} {2022})}\BibitemShut {NoStop}%
\bibitem [{\citenamefont {Milošević}\ and\ \citenamefont {Perali}(2015)}]{Milosevic_2015}%
  \BibitemOpen
  \bibfield  {author} {\bibinfo {author} {\bibfnamefont {M.~V.}\ \bibnamefont {Milošević}}\ and\ \bibinfo {author} {\bibfnamefont {A.}~\bibnamefont {Perali}},\ }\href {https://doi.org/10.1088/0953-2048/28/6/060201} {\bibfield  {journal} {\bibinfo  {journal} {Superconductor Science and Technology}\ }\textbf {\bibinfo {volume} {28}},\ \bibinfo {pages} {060201} (\bibinfo {year} {2015})}\BibitemShut {NoStop}%
\bibitem [{\citenamefont {Chubukov}\ \emph {et~al.}(2016)\citenamefont {Chubukov}, \citenamefont {Eremin},\ and\ \citenamefont {Efremov}}]{PhysRevB.93.174516}%
  \BibitemOpen
  \bibfield  {author} {\bibinfo {author} {\bibfnamefont {A.~V.}\ \bibnamefont {Chubukov}}, \bibinfo {author} {\bibfnamefont {I.}~\bibnamefont {Eremin}},\ and\ \bibinfo {author} {\bibfnamefont {D.~V.}\ \bibnamefont {Efremov}},\ }\href {https://doi.org/10.1103/PhysRevB.93.174516} {\bibfield  {journal} {\bibinfo  {journal} {Phys. Rev. B}\ }\textbf {\bibinfo {volume} {93}},\ \bibinfo {pages} {174516} (\bibinfo {year} {2016})}\BibitemShut {NoStop}%
\bibitem [{\citenamefont {Salasnich}\ \emph {et~al.}(2019)\citenamefont {Salasnich}, \citenamefont {Shanenko}, \citenamefont {Vagov}, \citenamefont {Aguiar},\ and\ \citenamefont {Perali}}]{PhysRevB.100.064510}%
  \BibitemOpen
  \bibfield  {author} {\bibinfo {author} {\bibfnamefont {L.}~\bibnamefont {Salasnich}}, \bibinfo {author} {\bibfnamefont {A.~A.}\ \bibnamefont {Shanenko}}, \bibinfo {author} {\bibfnamefont {A.}~\bibnamefont {Vagov}}, \bibinfo {author} {\bibfnamefont {J.~A.}\ \bibnamefont {Aguiar}},\ and\ \bibinfo {author} {\bibfnamefont {A.}~\bibnamefont {Perali}},\ }\href {https://doi.org/10.1103/PhysRevB.100.064510} {\bibfield  {journal} {\bibinfo  {journal} {Phys. Rev. B}\ }\textbf {\bibinfo {volume} {100}},\ \bibinfo {pages} {064510} (\bibinfo {year} {2019})}\BibitemShut {NoStop}%
\bibitem [{\citenamefont {Tajima}\ \emph {et~al.}(2019)\citenamefont {Tajima}, \citenamefont {Yerin}, \citenamefont {Perali},\ and\ \citenamefont {Pieri}}]{PhysRevB.99.180503}%
  \BibitemOpen
  \bibfield  {author} {\bibinfo {author} {\bibfnamefont {H.}~\bibnamefont {Tajima}}, \bibinfo {author} {\bibfnamefont {Y.}~\bibnamefont {Yerin}}, \bibinfo {author} {\bibfnamefont {A.}~\bibnamefont {Perali}},\ and\ \bibinfo {author} {\bibfnamefont {P.}~\bibnamefont {Pieri}},\ }\href {https://doi.org/10.1103/PhysRevB.99.180503} {\bibfield  {journal} {\bibinfo  {journal} {Phys. Rev. B}\ }\textbf {\bibinfo {volume} {99}},\ \bibinfo {pages} {180503} (\bibinfo {year} {2019})}\BibitemShut {NoStop}%
\bibitem [{\citenamefont {Tajima}\ \emph {et~al.}(2020)\citenamefont {Tajima}, \citenamefont {Yerin}, \citenamefont {Pieri},\ and\ \citenamefont {Perali}}]{PhysRevB.102.220504}%
  \BibitemOpen
  \bibfield  {author} {\bibinfo {author} {\bibfnamefont {H.}~\bibnamefont {Tajima}}, \bibinfo {author} {\bibfnamefont {Y.}~\bibnamefont {Yerin}}, \bibinfo {author} {\bibfnamefont {P.}~\bibnamefont {Pieri}},\ and\ \bibinfo {author} {\bibfnamefont {A.}~\bibnamefont {Perali}},\ }\href {https://doi.org/10.1103/PhysRevB.102.220504} {\bibfield  {journal} {\bibinfo  {journal} {Phys. Rev. B}\ }\textbf {\bibinfo {volume} {102}},\ \bibinfo {pages} {220504} (\bibinfo {year} {2020})}\BibitemShut {NoStop}%
\bibitem [{\citenamefont {Saraiva}\ \emph {et~al.}(2020)\citenamefont {Saraiva}, \citenamefont {Cavalcanti}, \citenamefont {Vagov}, \citenamefont {Vasenko}, \citenamefont {Perali}, \citenamefont {Dell'Anna},\ and\ \citenamefont {Shanenko}}]{PhysRevLett.125.217003}%
  \BibitemOpen
  \bibfield  {author} {\bibinfo {author} {\bibfnamefont {T.~T.}\ \bibnamefont {Saraiva}}, \bibinfo {author} {\bibfnamefont {P.~J.~F.}\ \bibnamefont {Cavalcanti}}, \bibinfo {author} {\bibfnamefont {A.}~\bibnamefont {Vagov}}, \bibinfo {author} {\bibfnamefont {A.~S.}\ \bibnamefont {Vasenko}}, \bibinfo {author} {\bibfnamefont {A.}~\bibnamefont {Perali}}, \bibinfo {author} {\bibfnamefont {L.}~\bibnamefont {Dell'Anna}},\ and\ \bibinfo {author} {\bibfnamefont {A.~A.}\ \bibnamefont {Shanenko}},\ }\href {https://doi.org/10.1103/PhysRevLett.125.217003} {\bibfield  {journal} {\bibinfo  {journal} {Phys. Rev. Lett.}\ }\textbf {\bibinfo {volume} {125}},\ \bibinfo {pages} {217003} (\bibinfo {year} {2020})}\BibitemShut {NoStop}%
\bibitem [{\citenamefont {Hanaguri}\ \emph {et~al.}(2019)\citenamefont {Hanaguri}, \citenamefont {Kasahara}, \citenamefont {B\"oker}, \citenamefont {Eremin}, \citenamefont {Shibauchi},\ and\ \citenamefont {Matsuda}}]{PhysRevLett.122.077001}%
  \BibitemOpen
  \bibfield  {author} {\bibinfo {author} {\bibfnamefont {T.}~\bibnamefont {Hanaguri}}, \bibinfo {author} {\bibfnamefont {S.}~\bibnamefont {Kasahara}}, \bibinfo {author} {\bibfnamefont {J.}~\bibnamefont {B\"oker}}, \bibinfo {author} {\bibfnamefont {I.}~\bibnamefont {Eremin}}, \bibinfo {author} {\bibfnamefont {T.}~\bibnamefont {Shibauchi}},\ and\ \bibinfo {author} {\bibfnamefont {Y.}~\bibnamefont {Matsuda}},\ }\href {https://doi.org/10.1103/PhysRevLett.122.077001} {\bibfield  {journal} {\bibinfo  {journal} {Phys. Rev. Lett.}\ }\textbf {\bibinfo {volume} {122}},\ \bibinfo {pages} {077001} (\bibinfo {year} {2019})}\BibitemShut {NoStop}%
\bibitem [{\citenamefont {Perali}\ \emph {et~al.}(2002)\citenamefont {Perali}, \citenamefont {Pieri}, \citenamefont {Strinati},\ and\ \citenamefont {Castellani}}]{perali2002}%
  \BibitemOpen
  \bibfield  {author} {\bibinfo {author} {\bibfnamefont {A.}~\bibnamefont {Perali}}, \bibinfo {author} {\bibfnamefont {P.}~\bibnamefont {Pieri}}, \bibinfo {author} {\bibfnamefont {G.~C.}\ \bibnamefont {Strinati}},\ and\ \bibinfo {author} {\bibfnamefont {C.}~\bibnamefont {Castellani}},\ }\href {https://doi.org/10.1103/PhysRevB.66.024510} {\bibfield  {journal} {\bibinfo  {journal} {Phys. Rev. B}\ }\textbf {\bibinfo {volume} {66}},\ \bibinfo {pages} {024510} (\bibinfo {year} {2002})}\BibitemShut {NoStop}%
\bibitem [{\citenamefont {Chen}\ \emph {et~al.}(2005)\citenamefont {Chen}, \citenamefont {Stajic}, \citenamefont {Tan},\ and\ \citenamefont {Levin}}]{chen2005bcs}%
  \BibitemOpen
  \bibfield  {author} {\bibinfo {author} {\bibfnamefont {Q.}~\bibnamefont {Chen}}, \bibinfo {author} {\bibfnamefont {J.}~\bibnamefont {Stajic}}, \bibinfo {author} {\bibfnamefont {S.}~\bibnamefont {Tan}},\ and\ \bibinfo {author} {\bibfnamefont {K.}~\bibnamefont {Levin}},\ }\href@noop {} {\bibfield  {journal} {\bibinfo  {journal} {Physics Reports}\ }\textbf {\bibinfo {volume} {412}},\ \bibinfo {pages} {1} (\bibinfo {year} {2005})}\BibitemShut {NoStop}%
\bibitem [{\citenamefont {Mueller}(2017)}]{mueller2017review}%
  \BibitemOpen
  \bibfield  {author} {\bibinfo {author} {\bibfnamefont {E.~J.}\ \bibnamefont {Mueller}},\ }\href@noop {} {\bibfield  {journal} {\bibinfo  {journal} {Reports on Progress in Physics}\ }\textbf {\bibinfo {volume} {80}},\ \bibinfo {pages} {104401} (\bibinfo {year} {2017})}\BibitemShut {NoStop}%
\bibitem [{\citenamefont {Strinati}\ \emph {et~al.}(2018)\citenamefont {Strinati}, \citenamefont {Pieri}, \citenamefont {R{\"o}pke}, \citenamefont {Schuck},\ and\ \citenamefont {Urban}}]{strinati2018bcs}%
  \BibitemOpen
  \bibfield  {author} {\bibinfo {author} {\bibfnamefont {G.~C.}\ \bibnamefont {Strinati}}, \bibinfo {author} {\bibfnamefont {P.}~\bibnamefont {Pieri}}, \bibinfo {author} {\bibfnamefont {G.}~\bibnamefont {R{\"o}pke}}, \bibinfo {author} {\bibfnamefont {P.}~\bibnamefont {Schuck}},\ and\ \bibinfo {author} {\bibfnamefont {M.}~\bibnamefont {Urban}},\ }\href@noop {} {\bibfield  {journal} {\bibinfo  {journal} {Phys. Rep.}\ }\textbf {\bibinfo {volume} {738}},\ \bibinfo {pages} {1} (\bibinfo {year} {2018})}\BibitemShut {NoStop}%
\bibitem [{\citenamefont {Ohashi}\ \emph {et~al.}(2020)\citenamefont {Ohashi}, \citenamefont {Tajima},\ and\ \citenamefont {{van Wyk}}}]{OHASHI2020103739}%
  \BibitemOpen
  \bibfield  {author} {\bibinfo {author} {\bibfnamefont {Y.}~\bibnamefont {Ohashi}}, \bibinfo {author} {\bibfnamefont {H.}~\bibnamefont {Tajima}},\ and\ \bibinfo {author} {\bibfnamefont {P.}~\bibnamefont {{van Wyk}}},\ }\href {https://doi.org/https://doi.org/10.1016/j.ppnp.2019.103739} {\bibfield  {journal} {\bibinfo  {journal} {Prog. Part. Nucl. Phys.}\ }\textbf {\bibinfo {volume} {111}},\ \bibinfo {pages} {103739} (\bibinfo {year} {2020})}\BibitemShut {NoStop}%
\bibitem [{\citenamefont {Innocenti}\ \emph {et~al.}(2010)\citenamefont {Innocenti}, \citenamefont {Poccia}, \citenamefont {Ricci}, \citenamefont {Valletta}, \citenamefont {Caprara}, \citenamefont {Perali},\ and\ \citenamefont {Bianconi}}]{PhysRevB.82.184528}%
  \BibitemOpen
  \bibfield  {author} {\bibinfo {author} {\bibfnamefont {D.}~\bibnamefont {Innocenti}}, \bibinfo {author} {\bibfnamefont {N.}~\bibnamefont {Poccia}}, \bibinfo {author} {\bibfnamefont {A.}~\bibnamefont {Ricci}}, \bibinfo {author} {\bibfnamefont {A.}~\bibnamefont {Valletta}}, \bibinfo {author} {\bibfnamefont {S.}~\bibnamefont {Caprara}}, \bibinfo {author} {\bibfnamefont {A.}~\bibnamefont {Perali}},\ and\ \bibinfo {author} {\bibfnamefont {A.}~\bibnamefont {Bianconi}},\ }\href {https://doi.org/10.1103/PhysRevB.82.184528} {\bibfield  {journal} {\bibinfo  {journal} {Phys. Rev. B}\ }\textbf {\bibinfo {volume} {82}},\ \bibinfo {pages} {184528} (\bibinfo {year} {2010})}\BibitemShut {NoStop}%
\bibitem [{\citenamefont {Suhl}\ \emph {et~al.}(1959)\citenamefont {Suhl}, \citenamefont {Matthias},\ and\ \citenamefont {Walker}}]{PhysRevLett.3.552}%
  \BibitemOpen
  \bibfield  {author} {\bibinfo {author} {\bibfnamefont {H.}~\bibnamefont {Suhl}}, \bibinfo {author} {\bibfnamefont {B.~T.}\ \bibnamefont {Matthias}},\ and\ \bibinfo {author} {\bibfnamefont {L.~R.}\ \bibnamefont {Walker}},\ }\href {https://doi.org/10.1103/PhysRevLett.3.552} {\bibfield  {journal} {\bibinfo  {journal} {Phys. Rev. Lett.}\ }\textbf {\bibinfo {volume} {3}},\ \bibinfo {pages} {552} (\bibinfo {year} {1959})}\BibitemShut {NoStop}%
\bibitem [{\citenamefont {Kondo}(1963)}]{10.1143/PTP.29.1}%
  \BibitemOpen
  \bibfield  {author} {\bibinfo {author} {\bibfnamefont {J.}~\bibnamefont {Kondo}},\ }\href {https://doi.org/10.1143/PTP.29.1} {\bibfield  {journal} {\bibinfo  {journal} {Progress of Theoretical Physics}\ }\textbf {\bibinfo {volume} {29}},\ \bibinfo {pages} {1} (\bibinfo {year} {1963})}\BibitemShut {NoStop}%
\bibitem [{\citenamefont {Nishiguchi}\ \emph {et~al.}(2013)\citenamefont {Nishiguchi}, \citenamefont {Kuroki}, \citenamefont {Arita}, \citenamefont {Oka},\ and\ \citenamefont {Aoki}}]{PhysRevB.88.014509}%
  \BibitemOpen
  \bibfield  {author} {\bibinfo {author} {\bibfnamefont {K.}~\bibnamefont {Nishiguchi}}, \bibinfo {author} {\bibfnamefont {K.}~\bibnamefont {Kuroki}}, \bibinfo {author} {\bibfnamefont {R.}~\bibnamefont {Arita}}, \bibinfo {author} {\bibfnamefont {T.}~\bibnamefont {Oka}},\ and\ \bibinfo {author} {\bibfnamefont {H.}~\bibnamefont {Aoki}},\ }\href {https://doi.org/10.1103/PhysRevB.88.014509} {\bibfield  {journal} {\bibinfo  {journal} {Phys. Rev. B}\ }\textbf {\bibinfo {volume} {88}},\ \bibinfo {pages} {014509} (\bibinfo {year} {2013})}\BibitemShut {NoStop}%
\bibitem [{\citenamefont {Yue}\ \emph {et~al.}(2022)\citenamefont {Yue}, \citenamefont {Aoki},\ and\ \citenamefont {Werner}}]{PhysRevB.106.L180506}%
  \BibitemOpen
  \bibfield  {author} {\bibinfo {author} {\bibfnamefont {C.}~\bibnamefont {Yue}}, \bibinfo {author} {\bibfnamefont {H.}~\bibnamefont {Aoki}},\ and\ \bibinfo {author} {\bibfnamefont {P.}~\bibnamefont {Werner}},\ }\href {https://doi.org/10.1103/PhysRevB.106.L180506} {\bibfield  {journal} {\bibinfo  {journal} {Phys. Rev. B}\ }\textbf {\bibinfo {volume} {106}},\ \bibinfo {pages} {L180506} (\bibinfo {year} {2022})}\BibitemShut {NoStop}%
\bibitem [{\citenamefont {Ochi}\ \emph {et~al.}(2022)\citenamefont {Ochi}, \citenamefont {Tajima}, \citenamefont {Iida},\ and\ \citenamefont {Aoki}}]{PhysRevResearch.4.013032}%
  \BibitemOpen
  \bibfield  {author} {\bibinfo {author} {\bibfnamefont {K.}~\bibnamefont {Ochi}}, \bibinfo {author} {\bibfnamefont {H.}~\bibnamefont {Tajima}}, \bibinfo {author} {\bibfnamefont {K.}~\bibnamefont {Iida}},\ and\ \bibinfo {author} {\bibfnamefont {H.}~\bibnamefont {Aoki}},\ }\href {https://doi.org/10.1103/PhysRevResearch.4.013032} {\bibfield  {journal} {\bibinfo  {journal} {Phys. Rev. Res.}\ }\textbf {\bibinfo {volume} {4}},\ \bibinfo {pages} {013032} (\bibinfo {year} {2022})}\BibitemShut {NoStop}%
\bibitem [{\citenamefont {Iskin}\ and\ \citenamefont {S\'a~de Melo}(2006)}]{PhysRevB.74.144517}%
  \BibitemOpen
  \bibfield  {author} {\bibinfo {author} {\bibfnamefont {M.}~\bibnamefont {Iskin}}\ and\ \bibinfo {author} {\bibfnamefont {C.~A.~R.}\ \bibnamefont {S\'a~de Melo}},\ }\href {https://doi.org/10.1103/PhysRevB.74.144517} {\bibfield  {journal} {\bibinfo  {journal} {Phys. Rev. B}\ }\textbf {\bibinfo {volume} {74}},\ \bibinfo {pages} {144517} (\bibinfo {year} {2006})}\BibitemShut {NoStop}%
\bibitem [{\citenamefont {Silaev}\ and\ \citenamefont {Babaev}(2011)}]{PhysRevB.84.094515}%
  \BibitemOpen
  \bibfield  {author} {\bibinfo {author} {\bibfnamefont {M.}~\bibnamefont {Silaev}}\ and\ \bibinfo {author} {\bibfnamefont {E.}~\bibnamefont {Babaev}},\ }\href {https://doi.org/10.1103/PhysRevB.84.094515} {\bibfield  {journal} {\bibinfo  {journal} {Phys. Rev. B}\ }\textbf {\bibinfo {volume} {84}},\ \bibinfo {pages} {094515} (\bibinfo {year} {2011})}\BibitemShut {NoStop}%
\bibitem [{\citenamefont {Silaev}\ and\ \citenamefont {Babaev}(2012)}]{PhysRevB.85.134514}%
  \BibitemOpen
  \bibfield  {author} {\bibinfo {author} {\bibfnamefont {M.}~\bibnamefont {Silaev}}\ and\ \bibinfo {author} {\bibfnamefont {E.}~\bibnamefont {Babaev}},\ }\href {https://doi.org/10.1103/PhysRevB.85.134514} {\bibfield  {journal} {\bibinfo  {journal} {Phys. Rev. B}\ }\textbf {\bibinfo {volume} {85}},\ \bibinfo {pages} {134514} (\bibinfo {year} {2012})}\BibitemShut {NoStop}%
\bibitem [{\citenamefont {Komendov\'a}\ \emph {et~al.}(2012)\citenamefont {Komendov\'a}, \citenamefont {Chen}, \citenamefont {Shanenko}, \citenamefont {Milo\ifmmode \check{s}\else \v{s}\fi{}evi\ifmmode~\acute{c}\else \'{c}\fi{}},\ and\ \citenamefont {Peeters}}]{PhysRevLett.108.207002}%
  \BibitemOpen
  \bibfield  {author} {\bibinfo {author} {\bibfnamefont {L.}~\bibnamefont {Komendov\'a}}, \bibinfo {author} {\bibfnamefont {Y.}~\bibnamefont {Chen}}, \bibinfo {author} {\bibfnamefont {A.~A.}\ \bibnamefont {Shanenko}}, \bibinfo {author} {\bibfnamefont {M.~V.}\ \bibnamefont {Milo\ifmmode \check{s}\else \v{s}\fi{}evi\ifmmode~\acute{c}\else \'{c}\fi{}}},\ and\ \bibinfo {author} {\bibfnamefont {F.~M.}\ \bibnamefont {Peeters}},\ }\href {https://doi.org/10.1103/PhysRevLett.108.207002} {\bibfield  {journal} {\bibinfo  {journal} {Phys. Rev. Lett.}\ }\textbf {\bibinfo {volume} {108}},\ \bibinfo {pages} {207002} (\bibinfo {year} {2012})}\BibitemShut {NoStop}%
\bibitem [{\citenamefont {Yerin}\ \emph {et~al.}(2019)\citenamefont {Yerin}, \citenamefont {Tajima}, \citenamefont {Pieri},\ and\ \citenamefont {Perali}}]{PhysRevB.100.104528}%
  \BibitemOpen
  \bibfield  {author} {\bibinfo {author} {\bibfnamefont {Y.}~\bibnamefont {Yerin}}, \bibinfo {author} {\bibfnamefont {H.}~\bibnamefont {Tajima}}, \bibinfo {author} {\bibfnamefont {P.}~\bibnamefont {Pieri}},\ and\ \bibinfo {author} {\bibfnamefont {A.}~\bibnamefont {Perali}},\ }\href {https://doi.org/10.1103/PhysRevB.100.104528} {\bibfield  {journal} {\bibinfo  {journal} {Phys. Rev. B}\ }\textbf {\bibinfo {volume} {100}},\ \bibinfo {pages} {104528} (\bibinfo {year} {2019})}\BibitemShut {NoStop}%
\bibitem [{\citenamefont {Midei}\ and\ \citenamefont {Perali}(2023)}]{PhysRevB.107.184501}%
  \BibitemOpen
  \bibfield  {author} {\bibinfo {author} {\bibfnamefont {G.}~\bibnamefont {Midei}}\ and\ \bibinfo {author} {\bibfnamefont {A.}~\bibnamefont {Perali}},\ }\href {https://doi.org/10.1103/PhysRevB.107.184501} {\bibfield  {journal} {\bibinfo  {journal} {Phys. Rev. B}\ }\textbf {\bibinfo {volume} {107}},\ \bibinfo {pages} {184501} (\bibinfo {year} {2023})}\BibitemShut {NoStop}%
\bibitem [{\citenamefont {Aoki}(2020)}]{aoki2020theoretical}%
  \BibitemOpen
  \bibfield  {author} {\bibinfo {author} {\bibfnamefont {H.}~\bibnamefont {Aoki}},\ }\href@noop {} {\bibfield  {journal} {\bibinfo  {journal} {Journal of Superconductivity and Novel Magnetism}\ }\textbf {\bibinfo {volume} {33}},\ \bibinfo {pages} {2341} (\bibinfo {year} {2020})}\BibitemShut {NoStop}%
\bibitem [{\citenamefont {Mazziotti}\ \emph {et~al.}(2021)\citenamefont {Mazziotti}, \citenamefont {Valletta}, \citenamefont {Raimondi},\ and\ \citenamefont {Bianconi}}]{PhysRevB.103.024523}%
  \BibitemOpen
  \bibfield  {author} {\bibinfo {author} {\bibfnamefont {M.~V.}\ \bibnamefont {Mazziotti}}, \bibinfo {author} {\bibfnamefont {A.}~\bibnamefont {Valletta}}, \bibinfo {author} {\bibfnamefont {R.}~\bibnamefont {Raimondi}},\ and\ \bibinfo {author} {\bibfnamefont {A.}~\bibnamefont {Bianconi}},\ }\href {https://doi.org/10.1103/PhysRevB.103.024523} {\bibfield  {journal} {\bibinfo  {journal} {Phys. Rev. B}\ }\textbf {\bibinfo {volume} {103}},\ \bibinfo {pages} {024523} (\bibinfo {year} {2021})}\BibitemShut {NoStop}%
\bibitem [{\citenamefont {Mazziotti}\ \emph {et~al.}(2022)\citenamefont {Mazziotti}, \citenamefont {Bianconi}, \citenamefont {Raimondi}, \citenamefont {Campi},\ and\ \citenamefont {Valletta}}]{10.1063/5.0123429}%
  \BibitemOpen
  \bibfield  {author} {\bibinfo {author} {\bibfnamefont {M.~V.}\ \bibnamefont {Mazziotti}}, \bibinfo {author} {\bibfnamefont {A.}~\bibnamefont {Bianconi}}, \bibinfo {author} {\bibfnamefont {R.}~\bibnamefont {Raimondi}}, \bibinfo {author} {\bibfnamefont {G.}~\bibnamefont {Campi}},\ and\ \bibinfo {author} {\bibfnamefont {A.}~\bibnamefont {Valletta}},\ }\href {https://doi.org/10.1063/5.0123429} {\bibfield  {journal} {\bibinfo  {journal} {Journal of Applied Physics}\ }\textbf {\bibinfo {volume} {132}},\ \bibinfo {pages} {193908} (\bibinfo {year} {2022})}\BibitemShut {NoStop}%
\bibitem [{\citenamefont {Logvenov}\ \emph {et~al.}(2023)\citenamefont {Logvenov}, \citenamefont {Bonmassar}, \citenamefont {Christiani}, \citenamefont {Campi}, \citenamefont {Valletta},\ and\ \citenamefont {Bianconi}}]{condmat8030078}%
  \BibitemOpen
  \bibfield  {author} {\bibinfo {author} {\bibfnamefont {G.}~\bibnamefont {Logvenov}}, \bibinfo {author} {\bibfnamefont {N.}~\bibnamefont {Bonmassar}}, \bibinfo {author} {\bibfnamefont {G.}~\bibnamefont {Christiani}}, \bibinfo {author} {\bibfnamefont {G.}~\bibnamefont {Campi}}, \bibinfo {author} {\bibfnamefont {A.}~\bibnamefont {Valletta}},\ and\ \bibinfo {author} {\bibfnamefont {A.}~\bibnamefont {Bianconi}},\ }\href {https://www.mdpi.com/2410-3896/8/3/78} {\bibfield  {journal} {\bibinfo  {journal} {Condens. Matter}\ }\textbf {\bibinfo {volume} {8}},\ \bibinfo {pages} {78} (\bibinfo {year} {2023})}\BibitemShut {NoStop}%
\bibitem [{\citenamefont {Gor’kov}\ and\ \citenamefont {Melik-Barkhudarov}(1961)}]{gor1961contribution}%
  \BibitemOpen
  \bibfield  {author} {\bibinfo {author} {\bibfnamefont {L.}~\bibnamefont {Gor’kov}}\ and\ \bibinfo {author} {\bibfnamefont {T.}~\bibnamefont {Melik-Barkhudarov}},\ }\href@noop {} {\bibfield  {journal} {\bibinfo  {journal} {Sov. Phys. JETP}\ }\textbf {\bibinfo {volume} {13}},\ \bibinfo {pages} {1018} (\bibinfo {year} {1961})}\BibitemShut {NoStop}%
\bibitem [{\citenamefont {Pisani}\ \emph {et~al.}(2018)\citenamefont {Pisani}, \citenamefont {Perali}, \citenamefont {Pieri},\ and\ \citenamefont {Strinati}}]{PhysRevB.97.014528}%
  \BibitemOpen
  \bibfield  {author} {\bibinfo {author} {\bibfnamefont {L.}~\bibnamefont {Pisani}}, \bibinfo {author} {\bibfnamefont {A.}~\bibnamefont {Perali}}, \bibinfo {author} {\bibfnamefont {P.}~\bibnamefont {Pieri}},\ and\ \bibinfo {author} {\bibfnamefont {G.~C.}\ \bibnamefont {Strinati}},\ }\href {https://doi.org/10.1103/PhysRevB.97.014528} {\bibfield  {journal} {\bibinfo  {journal} {Phys. Rev. B}\ }\textbf {\bibinfo {volume} {97}},\ \bibinfo {pages} {014528} (\bibinfo {year} {2018})}\BibitemShut {NoStop}%
\bibitem [{\citenamefont {Link}\ \emph {et~al.}(2023)\citenamefont {Link}, \citenamefont {Gao}, \citenamefont {Kell}, \citenamefont {Breyer}, \citenamefont {Eberz}, \citenamefont {Rauf},\ and\ \citenamefont {K\"ohl}}]{PhysRevLett.130.203401}%
  \BibitemOpen
  \bibfield  {author} {\bibinfo {author} {\bibfnamefont {M.}~\bibnamefont {Link}}, \bibinfo {author} {\bibfnamefont {K.}~\bibnamefont {Gao}}, \bibinfo {author} {\bibfnamefont {A.}~\bibnamefont {Kell}}, \bibinfo {author} {\bibfnamefont {M.}~\bibnamefont {Breyer}}, \bibinfo {author} {\bibfnamefont {D.}~\bibnamefont {Eberz}}, \bibinfo {author} {\bibfnamefont {B.}~\bibnamefont {Rauf}},\ and\ \bibinfo {author} {\bibfnamefont {M.}~\bibnamefont {K\"ohl}},\ }\href {https://doi.org/10.1103/PhysRevLett.130.203401} {\bibfield  {journal} {\bibinfo  {journal} {Phys. Rev. Lett.}\ }\textbf {\bibinfo {volume} {130}},\ \bibinfo {pages} {203401} (\bibinfo {year} {2023})}\BibitemShut {NoStop}%
\bibitem [{\citenamefont {Yu}\ \emph {et~al.}(2009)\citenamefont {Yu}, \citenamefont {Huang},\ and\ \citenamefont {Yin}}]{PhysRevA.79.053636}%
  \BibitemOpen
  \bibfield  {author} {\bibinfo {author} {\bibfnamefont {Z.-Q.}\ \bibnamefont {Yu}}, \bibinfo {author} {\bibfnamefont {K.}~\bibnamefont {Huang}},\ and\ \bibinfo {author} {\bibfnamefont {L.}~\bibnamefont {Yin}},\ }\href {https://doi.org/10.1103/PhysRevA.79.053636} {\bibfield  {journal} {\bibinfo  {journal} {Phys. Rev. A}\ }\textbf {\bibinfo {volume} {79}},\ \bibinfo {pages} {053636} (\bibinfo {year} {2009})}\BibitemShut {NoStop}%
\bibitem [{\citenamefont {Thouless}(1960)}]{THOULESS1960553}%
  \BibitemOpen
  \bibfield  {author} {\bibinfo {author} {\bibfnamefont {D.~J.}\ \bibnamefont {Thouless}},\ }\href {https://doi.org/https://doi.org/10.1016/0003-4916(60)90122-6} {\bibfield  {journal} {\bibinfo  {journal} {Annals of Physics}\ }\textbf {\bibinfo {volume} {10}},\ \bibinfo {pages} {553} (\bibinfo {year} {1960})}\BibitemShut {NoStop}%
\bibitem [{Sup()}]{Supplement}%
  \BibitemOpen
  \href@noop {} {}\bibinfo {howpublished} {See the Supplemental Material for details of the theoretical framework and additional numerical results for different values of parameters.}\BibitemShut {Stop}%
\bibitem [{\citenamefont {Aoki}\ and\ \citenamefont {Kuroki}(1990)}]{PhysRevB.42.2125}%
  \BibitemOpen
  \bibfield  {author} {\bibinfo {author} {\bibfnamefont {H.}~\bibnamefont {Aoki}}\ and\ \bibinfo {author} {\bibfnamefont {K.}~\bibnamefont {Kuroki}},\ }\href {https://doi.org/10.1103/PhysRevB.42.2125} {\bibfield  {journal} {\bibinfo  {journal} {Phys. Rev. B}\ }\textbf {\bibinfo {volume} {42}},\ \bibinfo {pages} {2125} (\bibinfo {year} {1990})}\BibitemShut {NoStop}%
\bibitem [{\citenamefont {Kuroki}\ and\ \citenamefont {Aoki}(1992)}]{PhysRevLett.69.3820}%
  \BibitemOpen
  \bibfield  {author} {\bibinfo {author} {\bibfnamefont {K.}~\bibnamefont {Kuroki}}\ and\ \bibinfo {author} {\bibfnamefont {H.}~\bibnamefont {Aoki}},\ }\href {https://doi.org/10.1103/PhysRevLett.69.3820} {\bibfield  {journal} {\bibinfo  {journal} {Phys. Rev. Lett.}\ }\textbf {\bibinfo {volume} {69}},\ \bibinfo {pages} {3820} (\bibinfo {year} {1992})}\BibitemShut {NoStop}%
\bibitem [{\citenamefont {Kuroki}\ and\ \citenamefont {Aoki}(1993)}]{PhysRevB.48.7598}%
  \BibitemOpen
  \bibfield  {author} {\bibinfo {author} {\bibfnamefont {K.}~\bibnamefont {Kuroki}}\ and\ \bibinfo {author} {\bibfnamefont {H.}~\bibnamefont {Aoki}},\ }\href {https://doi.org/10.1103/PhysRevB.48.7598} {\bibfield  {journal} {\bibinfo  {journal} {Phys. Rev. B}\ }\textbf {\bibinfo {volume} {48}},\ \bibinfo {pages} {7598} (\bibinfo {year} {1993})}\BibitemShut {NoStop}%
\bibitem [{not()}]{note}%
  \BibitemOpen
  \href@noop {} {}\bibinfo {howpublished} {Mixed particle-particle and particle-hole diagrams have been evoked in a different context and formulated as the dynamical vertex approximation for repulsive lattice models in M. Kitatani, T. Sch\"afer, H. Aoki, and K. Held, \href{https://link.aps.org/doi/10.1103/PhysRevB.99.041115}{Phys. Rev. B \textbf{99}, 041115(R) (2019).}}\BibitemShut {Stop}%
\bibitem [{\citenamefont {Sayyad}\ \emph {et~al.}(2020)\citenamefont {Sayyad}, \citenamefont {Huang}, \citenamefont {Kitatani}, \citenamefont {Vaezi}, \citenamefont {Nussinov}, \citenamefont {Vaezi},\ and\ \citenamefont {Aoki}}]{PhysRevB.101.014501}%
  \BibitemOpen
  \bibfield  {author} {\bibinfo {author} {\bibfnamefont {S.}~\bibnamefont {Sayyad}}, \bibinfo {author} {\bibfnamefont {E.~W.}\ \bibnamefont {Huang}}, \bibinfo {author} {\bibfnamefont {M.}~\bibnamefont {Kitatani}}, \bibinfo {author} {\bibfnamefont {M.-S.}\ \bibnamefont {Vaezi}}, \bibinfo {author} {\bibfnamefont {Z.}~\bibnamefont {Nussinov}}, \bibinfo {author} {\bibfnamefont {A.}~\bibnamefont {Vaezi}},\ and\ \bibinfo {author} {\bibfnamefont {H.}~\bibnamefont {Aoki}},\ }\href {https://doi.org/10.1103/PhysRevB.101.014501} {\bibfield  {journal} {\bibinfo  {journal} {Phys. Rev. B}\ }\textbf {\bibinfo {volume} {101}},\ \bibinfo {pages} {014501} (\bibinfo {year} {2020})}\BibitemShut {NoStop}%
\bibitem [{\citenamefont {Sayyad}\ \emph {et~al.}(2023)\citenamefont {Sayyad}, \citenamefont {Kitatani}, \citenamefont {Vaezi},\ and\ \citenamefont {Aoki}}]{Sayyad_2023}%
  \BibitemOpen
  \bibfield  {author} {\bibinfo {author} {\bibfnamefont {S.}~\bibnamefont {Sayyad}}, \bibinfo {author} {\bibfnamefont {M.}~\bibnamefont {Kitatani}}, \bibinfo {author} {\bibfnamefont {A.}~\bibnamefont {Vaezi}},\ and\ \bibinfo {author} {\bibfnamefont {H.}~\bibnamefont {Aoki}},\ }\href {https://doi.org/10.1088/1361-648X/acc6af} {\bibfield  {journal} {\bibinfo  {journal} {Journal of Physics: Condensed Matter}\ }\textbf {\bibinfo {volume} {35}},\ \bibinfo {pages} {245605} (\bibinfo {year} {2023})}\BibitemShut {NoStop}%
\bibitem [{\citenamefont {Yamazaki}\ \emph {et~al.}(2020)\citenamefont {Yamazaki}, \citenamefont {Ochi}, \citenamefont {Ogura}, \citenamefont {Kuroki}, \citenamefont {Eisaki}, \citenamefont {Uchida},\ and\ \citenamefont {Aoki}}]{PhysRevResearch.2.033356}%
  \BibitemOpen
  \bibfield  {author} {\bibinfo {author} {\bibfnamefont {K.}~\bibnamefont {Yamazaki}}, \bibinfo {author} {\bibfnamefont {M.}~\bibnamefont {Ochi}}, \bibinfo {author} {\bibfnamefont {D.}~\bibnamefont {Ogura}}, \bibinfo {author} {\bibfnamefont {K.}~\bibnamefont {Kuroki}}, \bibinfo {author} {\bibfnamefont {H.}~\bibnamefont {Eisaki}}, \bibinfo {author} {\bibfnamefont {S.}~\bibnamefont {Uchida}},\ and\ \bibinfo {author} {\bibfnamefont {H.}~\bibnamefont {Aoki}},\ }\href {https://doi.org/10.1103/PhysRevResearch.2.033356} {\bibfield  {journal} {\bibinfo  {journal} {Phys. Rev. Res.}\ }\textbf {\bibinfo {volume} {2}},\ \bibinfo {pages} {033356} (\bibinfo {year} {2020})}\BibitemShut {NoStop}%
\bibitem [{\citenamefont {Kosterlitz}(2016)}]{kosterlitz2016kosterlitz}%
  \BibitemOpen
  \bibfield  {author} {\bibinfo {author} {\bibfnamefont {J.~M.}\ \bibnamefont {Kosterlitz}},\ }\href@noop {} {\bibfield  {journal} {\bibinfo  {journal} {Reports on Progress in Physics}\ }\textbf {\bibinfo {volume} {79}},\ \bibinfo {pages} {026001} (\bibinfo {year} {2016})}\BibitemShut {NoStop}%
\bibitem [{\citenamefont {Midei}\ \emph {et~al.}(2024)\citenamefont {Midei}, \citenamefont {Furutani}, \citenamefont {Salasnich},\ and\ \citenamefont {Perali}}]{midei2024predictive}%
  \BibitemOpen
  \bibfield  {author} {\bibinfo {author} {\bibfnamefont {G.}~\bibnamefont {Midei}}, \bibinfo {author} {\bibfnamefont {K.}~\bibnamefont {Furutani}}, \bibinfo {author} {\bibfnamefont {L.}~\bibnamefont {Salasnich}},\ and\ \bibinfo {author} {\bibfnamefont {A.}~\bibnamefont {Perali}},\ }\href@noop {} {\bibfield  {journal} {\bibinfo  {journal} {arXiv preprint arXiv:2403.03025}\ } (\bibinfo {year} {2024})}\BibitemShut {NoStop}%
\bibitem [{\citenamefont {Paramasivam}\ \emph {et~al.}(2023)\citenamefont {Paramasivam}, \citenamefont {Gangadharan}, \citenamefont {Milo{\v{s}}evi{\'c}},\ and\ \citenamefont {Perali}}]{paramasivam2023high}%
  \BibitemOpen
  \bibfield  {author} {\bibinfo {author} {\bibfnamefont {S.~K.}\ \bibnamefont {Paramasivam}}, \bibinfo {author} {\bibfnamefont {S.~P.}\ \bibnamefont {Gangadharan}}, \bibinfo {author} {\bibfnamefont {M.~V.}\ \bibnamefont {Milo{\v{s}}evi{\'c}}},\ and\ \bibinfo {author} {\bibfnamefont {A.}~\bibnamefont {Perali}},\ }\href@noop {} {\bibfield  {journal} {\bibinfo  {journal} {arXiv preprint arXiv:2312.09017}\ } (\bibinfo {year} {2023})}\BibitemShut {NoStop}%
\end{thebibliography}%


\clearpage

\beginsupplement

\setcounter{equation}{0}
\setcounter{figure}{0}
\setcounter{table}{0}
\setcounter{page}{1}
\makeatletter
\renewcommand{\theequation}{S\arabic{equation}}
\renewcommand{\thefigure}{S\arabic{figure}}
\renewcommand{\bibnumfmt}[1]{[S#1]}

\section*{Supplemental Material}

\subsection{Two-band many-body $T$-matrix\\
with the GMB correction}
Here we formulate how we can implement the GMB theory to the two-band system that accommodates bothe of intraband and interband couplings based on the many-body $T$-matrix formalism.
Before introducing the GMB corrections,
let us first look at how to determine $T_{\rm c}$ from the infrared behavior of the two-band many-body $T$-matrix.
For that, we can define band-indexed matrix forms (hatted) for the intra- and inter-band couplings as
\begin{align} 
\hat{U}=
\left(\begin{array}{cc}
U_{11} & U_{12} \\
U_{21} & U_{22} 
\end{array}
\right),
\end{align}
and the bare pair susceptibility
\begin{align} 
\hat{\Pi}(q)=
\left(\begin{array}{cc}
\Pi_{11}(q) & 0 \\
0 & \Pi_{22}(q) 
\end{array}
\right),
\end{align}
with the four-component index
$q=(\bm{q},i\nu_\ell)$ where
$i\nu_\ell=2\ell\pi T$ ($\ell \in \mathbb{Z}$) is a bosonic Matsubara frequency.
$\Pi_{nn}(q)$ is given by
\begin{align}
\Pi_{nn}(q)&=T\sum_{\bm{k},i\omega_s}G_{n}(k+q)
G_{n}(-k)\cr
&=-\sum_{\bm{k}}\frac{1-f(\xi_{\bm{k}+\bm{q},n})-f(\xi_{-\bm{k},n})}{i\nu_\ell-\xi_{\bm{k}+\bm{q},n}-\xi_{-\bm{k},n}},
\end{align}
where $G_{n}(k)=1/(i\omega_\ell-\xi_{\bm{k},n})$ is the bare Green's function for band $n$, $i\omega_\ell=(2\ell+1)\pi T$ is a fermionic Matsubara frequency,
and $f(\xi)=\textcolor{blue}{1/}(e^{\xi/T}+1)$ is the Fermi-Dirac distribution function.
If we neglect the GMB correction, the many-body $T$-matrix reads
\begin{align}
    \hat{\Gamma}(q)=\left[1+\hat{U}\Pi(q)\right]^{-1}\hat{U}.
\end{align}
The Thouless criterion~\cite{THOULESS1960553} indicates that the superconducting critical temperature can be obtained from the condition~\cite{PhysRevB.99.180503,PhysRevB.102.220504}
\begin{align}
\label{eq:Thouless}
    {\rm det}\left[1+\hat{U}\hat{\Pi}(q=0)\right]=0,
\end{align}
which leads to the BCS-type critical temperature $T_{\rm c}^{\rm BCS}$, here for the two-band system.

In the two-band system with the intraband and interband couplings, 
we need to consider the screening effects on both couplings.
The GMB screening of the intraband interaction can be implemented straightforwardly by replacing the diagonal interactions with screened ones as
\begin{align}
\label{eqs:usc}
    \hat{U}^{\rm sc}(q)=\left(
    \begin{array}{cc}
      U_{11}^{\rm sc}(q)   & U_{12}  \\
      U_{21}   & U_{22}^{\rm sc}(q) 
    \end{array}
    \right),
\end{align}
with
\begin{align}
    U_{nn}^{\rm sc}(q)=
    \frac{U_{nn}}{1+U_{nn}\chi_{nn}(q)},
\end{align}
where
\begin{align}
\chi_{nn}(q)&=T\sum_{\bm{k},i\omega_k}G_{n}(k+q)G_{n}(k)\cr
&=\sum_{\bm{k}}\frac{f(\xi_{\bm{k},n})-f(\xi_{\bm{k}+\bm{q},n})}{i\nu_\ell
    +\xi_{\bm{k},n}-\xi_{\bm{k}+\bm{q},n}}
\end{align}
is the particle-hole bubble in band $n$. 
For simplicity, we employ a low-energy approximation around the Fermi surface as in Ref.~\cite{PhysRevA.79.053636}.
When $\mu-E_0\delta_{n,2}>0$, we obtain the averaged particle-hole bubble as
\begin{align}
\label{eqs:phb}
\left<\chi_{nn}\right>=\frac{m_n}{4\pi^2}{}\int_{-1}^{1}d\cos\theta\int_{0}^{\infty}\frac{kdk}{q_n}f(\xi_{\bm{k},n}){\rm ln}\left|\frac{q_n-2k}{q_n+2k}\right|,
\end{align}
where we have used $i\nu_\ell\simeq 0$ and $|\bm{q}|\simeq q_n\equiv \sqrt{2m_n(\mu-E_0\delta_{n,2})(1+\cos\theta)}$
with $\theta$ being the angle 
between incoming and outgoing momenta.  
When $\mu-E_0\delta_{n,2}<0$ where the Fermi surface does not exist for band $n$, we get
\begin{align}
\left<\chi_{nn}\right>=-\frac{m_n}{2\pi^2}\int_{0}^{\infty}dkf(\xi_{\bm{k},n}).
\end{align}
In this way,
we obtain the screened intraband coupling with the averaged particle-hole bubble as
\begin{align}
    U_{nn}^{\rm sc}(q)\simeq \frac{U_{nn}}{1+U_{nn}\langle \chi_{nn}\rangle }.
\end{align}
The many-body $T$-matrix with the screened intraband coupling then reads
\begin{align}
\label{eq:tmat1}
    \hat{\Gamma}(q)&=\hat{U}^{\rm sc}(q)-
    \hat{U}^{\rm sc}(q)\hat{\Pi}(q)\hat{\Gamma}(q)\cr
    &=\left[1+\hat{U}^{\rm sc}(q)\hat{\Pi}(q)\right]^{-1}\hat{U}^{\rm sc}(q).
\end{align}

Let us now turn to the screening correction for the interband pair-exchange coupling.  
While the screening correction for the \textit{intraband} coupling is included in Eq.~\eqref{eq:tmat1}, that for \textit{interband} one is not. 
So let us include the screening correction.
To this end, we introduce the effective intraband coupling induced by the interband pair-exchange process, which we can call the Fano-Feshbach-like, as
\begin{align}
    \hat{\lambda}(q)
    &\equiv
    \left(
    \begin{array}{cc}
       \lambda_{11}(q)  & 0  \\
        0 & \lambda_{22}(q)
    \end{array}
    \right)\cr
    &=
    \left(
    \begin{array}{cc}
       -U_{12}\Pi_{22}(q)U_{21}  & 0  \\
        0 & -U_{21}\Pi_{11}(q)U_{12}
    \end{array}
    \right).
\end{align}
The screening effect on $\hat{\lambda}(q)$
should be taken into account 
as in the case of the intraband attractions, 
so let us include the screening effect on $\hat{\lambda}(q)$ in the many-body $T$-matrix by summing up the diagrams in Fig.~1(b)
in the main text as
\begin{align}
\label{eqs:gamma}
    \hat{\Gamma}(q)
    &=
    \hat{U}^{\rm sc}(q)-
    \hat{U}^{\rm sc}(q)\hat{\Pi}(q)\hat{\Gamma}(q)
    -\hat{\lambda}(q)\langle \hat{\chi}\rangle
    \hat{\Gamma}_{\rm d}(q),
\end{align}
where we have introduced the diagonal particle-hole bubble $\langle\hat{\chi}\rangle ={\rm diag}(\langle\chi_{11}\rangle,\langle\chi_{22}\rangle)$, along with the diagonal $T$-matrix $\hat{\Gamma}_{\rm d}(q)={\rm diag}\left(\Gamma_{11}(q),\Gamma_{22}(q)\right)$.

If one wants to single out, in Eq.~\eqref{eqs:gamma}, how the pair-exchange induced coupling $\hat{\lambda}(q)$ is screened, 
one can put $U_{11}=U_{22}=0$, which yields 
\begin{align}
    \Gamma_{nn}(q)
    &=\frac{\lambda_{nn}^{\rm sc}(q)}{1+\lambda_{nn}^{\rm sc}(q)\Pi_{nn}(q)},
\label{eq:gammann}
\end{align}
with the screened effective interaction defined as
\begin{align}
    \lambda_{nn}^{\rm sc}(q)=\frac{\lambda_{nn}(q)}{1+\lambda_{nn}(q)\langle\chi_{nn}\rangle}.
\end{align}
Equation~\eqref{eq:gammann} reduces to the single-band case if $\lambda_{nn}$, $\Pi_{nn}$, and $\langle\chi_{nn}\rangle$ are replaced with the single-band counterparts~\cite{PhysRevA.79.053636}.
This confirms that our scheme correctly reproduces the GMB correction with respect to the pair-exchange-induced interactions.

\subsection{Derivation of Eq. (6) 
in the main text}
In the absence of the GMB correction, the Thouless criterion given by Eq.~\eqref{eq:Thouless} reads
\begin{align}
\label{eq:Thouless2}
    &[1+U_{11}\Pi_{11}(0)][1+U_{22}\Pi_{22}(0)]\cr
    &-U_{12}U_{21}\Pi_{11}(0)\Pi_{22}(0)=0.
\end{align}
Dividing the left and right hand sides of Eq.~\eqref{eq:Thouless2} by $1+U_{11}\Pi_{11}(0)$ (which is nonzero because of weak $U_{11}$), we find
\begin{align}
\label{eq:Thouless3}
    1+\left[U_{22}-\frac{U_{12}U_{21}\Pi_{11}(0)}{1+U_{11}\Pi_{11}(0)}\right]\Pi_{22}(0)=0.
\end{align}
For $m_2\rightarrow \infty$ (i.e., $\xi_{\bm{k},2}\rightarrow E_0-\mu$), we end up with
\begin{align}
\label{eq:pi22_largemass}
    \Pi_{22}(0)&=\frac{1}{(2\pi)^3}\int_0^{\Lambda}k^2dk\frac{1-2f(\xi_{\bm{k},2})}{2\xi_{\bm{k},2}}\cr
    &\rightarrow
    \frac{\Lambda^3}{6\pi^2}\mathcal{F}(E_0-\mu) \quad (m_2\rightarrow \infty),
\end{align}
where $\mathcal{F}(x)=\frac{\tanh\left(\frac{x}{2T_{\rm c}}\right)}{2x}$.
Substituting Eq.~\eqref{eq:pi22_largemass} into Eq.~\eqref{eq:Thouless3}, we obtain Eq.~ (6).

\subsection{Numerical calculation of the many-body $T$-matrix with GMB corrections}
In contrast to Eq.~\eqref{eq:Thouless},
one needs to perform the successive substitution 
to obtain $\hat{\Gamma}(q=0)$ in the 
screening-corrected Eq.~\eqref{eqs:gamma} 
because the many-body $T$-matrix is no longer in a closed form
due to the GMB correction for the interband coupling.
The lowest-order contribution at $q=0$ reads
\begin{align}
\left(\begin{array}{cc}
       \Gamma_{11}^{(0)}  & \Gamma_{12}^{(0)} \\
       \Gamma_{21}^{(0)}  & \Gamma_{22}^{(0)} 
    \end{array}\right)
    =
    \left(\begin{array}{cc}
       U_{11}^{\rm sc}  & U_{12} \\
       U_{21}  & U_{22}^{\rm sc} 
    \end{array}\right),
\end{align}
for $\Gamma_{nn'}^{(j)}$ where $j$ is the iteration number 
and the argument $(q=0)$ is omitted.
From Eq.~(\ref{eqs:gamma}),
one can obtain the recurrence relation for $\Gamma_{nn'}^{(j)}$ as
\begin{widetext}    
\begin{align}
\label{eq:iteration}
    &\left(\begin{array}{cc}
       \Gamma_{11}^{(j+1)}  & \Gamma_{12}^{(j+1)} \\
       \Gamma_{21}^{(j+1)}  & \Gamma_{22}^{(j+1)} 
    \end{array}\right)
    =
        \left(
\begin{array}{cc}
 U_{11}^{\rm sc}-[U_{11}^{\rm sc}\Pi_{11}+\lambda_{11}\langle \chi_{11}\rangle]\Gamma_{11}^{(j)} -U_{12}\Pi_{22}\Gamma_{21}^{(j)}&
 U_{12}-U_{11}^{\rm sc}\Pi_{11}\Gamma_{12}^{(j)}-U_{12}\Pi_{22}\Gamma_{22}^{(j)} \\
 U_{21}-U_{21}\Pi_{11}\Gamma_{11}^{(j)}-U_{22}^{\rm sc}\Pi_{22}\Gamma_{21}^{(j)}
 & U_{22}^{\rm sc}-[U_{22}^{\rm sc}\Pi_{22}+\lambda_{22}\langle \chi_{22}\rangle]
 \Gamma_{22}^{(j)}-U_{21}\Pi_{11}\Gamma_{12}^{(j)}
\end{array}\right).
\end{align}
\end{widetext}    
In this work, we have calculate $\Gamma_{nn'}^{(j)}$ up to large enough $j$ ($=10000$ here) to determine the critical temperature $T_{\rm c}$, defined here as the temperature at which large enough $\Gamma_{12}/U_{12}$ ($\Gamma_{12}/U_{12}=10^6$ here) is achieved.  
According to the Thouless criterion, all the $T$-matrices $\Gamma_{11}$, $\Gamma_{12}$, $\Gamma_{21}$, and $\Gamma_{22}$ diverge as shown in Eqs. (S4) and (S5).
Thus a blown-up $\Gamma_{12}$ can be taken as a signature of approached $T_{\rm c}$, which is shown to be robust against 
changes of the number of iterations.
We have also confirmed that the calculation with the above scheme accurately reproduces Eq.~\eqref{eq:Thouless} when the GMB correction is switched off.

\subsection{Fano-Feshbach resonance in the heavy band}
\label{app:FFresonance}
While we take the band-independent intraband couplings $U_{22}=U_{11}$ 
for simplicity, we can expect the heavy effective mass in band 2 may induce two-body bound or resonant states even in that case.
When the interband interaction is absent ($U_{12}=U_{21}=0$),
the system can indeed have a bound state in the energy continuum.
The two-body binding energy $E_{\rm b,2}$ measured from the bottom of the heavy band $2E_0$ satisfies
\begin{align}
    1+U_{22}\Pi_{22}^{\rm vac}(\omega=2E_0-E_{\rm b,2})=0,
\end{align}
where 
\begin{align}
\label{eq:pi_vac}
    \Pi_{nn}^{\rm vac}(\omega)=\sum_{\bm{k}}\frac{1}{\omega-k^2/m_n-2E_0\delta_{n,2}}
\end{align}
is the two-particle propagator without medium effects.
After the momentum summation in Eq.~\eqref{eq:pi_vac}, we find
\begin{align}
    1&=\frac{m_2U_{22}}{2\pi^2}
    \left[\sqrt{m_2E_{\rm b,2}}\tan^{-1}\left(\frac{\Lambda}{\sqrt{m_2E_{\rm b,2}}}\right)-\Lambda\right].
\end{align}
For a large cutoff (i.e., $\tan^{-1}\left({\Lambda}/\sqrt{m_2E_{\rm b,2}}\right)\rightarrow \frac{\pi}{2}$), we obtain
\begin{align}
    {{E_{\rm b,2}}}
    =\frac{4}{\pi^2 m_2}\left(\frac{2\pi^2}{m_2U_{11}}+{\Lambda}\right)^2\theta\left(\frac{2\pi^2}{m_2U_{11}}+{\Lambda}\right),
\end{align}
where $\theta(x)$ is the Heaviside step function. 
Such a bound state plays a crucial role in the superconducting transition near the incipient heavy band with weak interband couplings~\cite{PhysRevResearch.4.013032}.
Incidentally, increased effective-mass can also lead to the unitary limit, where the two-body bound state starts to appear at
\begin{align}
    \frac{m_2}{m_1}=1-\frac{\pi}{2a_{11}\Lambda},
\end{align}
which indicates that the result depends on the cutoff $\Lambda$.

In the presence of $U_{12}$, the bound state in the continuum turns into a 
Fano-Feshbach resonant state with a finite lifetime.
Namely, the pole of the two-body $T$-matrix locates in the complex energy plane (i.e., $\omega_{\rm res}\in \mathbb{C}$) as
\begin{align}
    1+U_{22}\Pi_{22}^{\rm vac}(\omega_{\rm res})-\frac{U_{12}^2\Pi_{11}^{\rm vac}(\omega_{\rm res})\Pi_{22}^{\rm vac}(\omega_{\rm res})}{1+U_{11}\Pi_{11}(\omega_{\rm res})}=0.
\end{align}
Considering the perturbation with respect to $U_{12}$, one may find the resonance energy as
\begin{align}
\label{eq:resonance}
        &\omega_{\rm res}\simeq -E_{\rm b,2}+2E_0\cr
        &-\frac{16\pi^2U_{12}^4}{m_2^3U_{22}^4}\frac{\left(\frac{m_1\Lambda}{2\pi^2}+i\frac{m_1\sqrt{2m_1E_0}}{4\pi}\right)^2}{\left[1+U_{11}\left(\frac{m_1\Lambda}{2\pi^2}+i\frac{m_1\sqrt{2m_1E_0}}{4\pi}\right)\right]^2}.
\end{align}
In this manner, the introduction of $U_{12}$ leads to a shift of $E_{\rm b,2}$ along with an imaginary part.  
The enlarged $E_{\rm b,2}$ can be regarded as the enhancement of two-body attraction due to the pair-exchange coupling.
In the limit of $\Lambda\rightarrow\infty$,
we analytically find that the magnitude of $E_{\rm b,2}$ is enlarged by $16\pi^2U_{12}^4\left[m_2^3U_{22}^4\left(\frac{2\pi^2}{m_1\Lambda}+U_{11}\right)^2\right]^{-1} \;(>0)$.
However, if $U_{12}$ is sufficiently small, this resonant state can approximately be regarded as a bound state with $\omega_{\rm res}\simeq -E_{\rm b,2}+2E_0$.

\subsection{Superconducting critical temperature for different parameters}
\label{app:1}

 \begin{figure}[t]
    \centering
    \includegraphics[width=8cm]{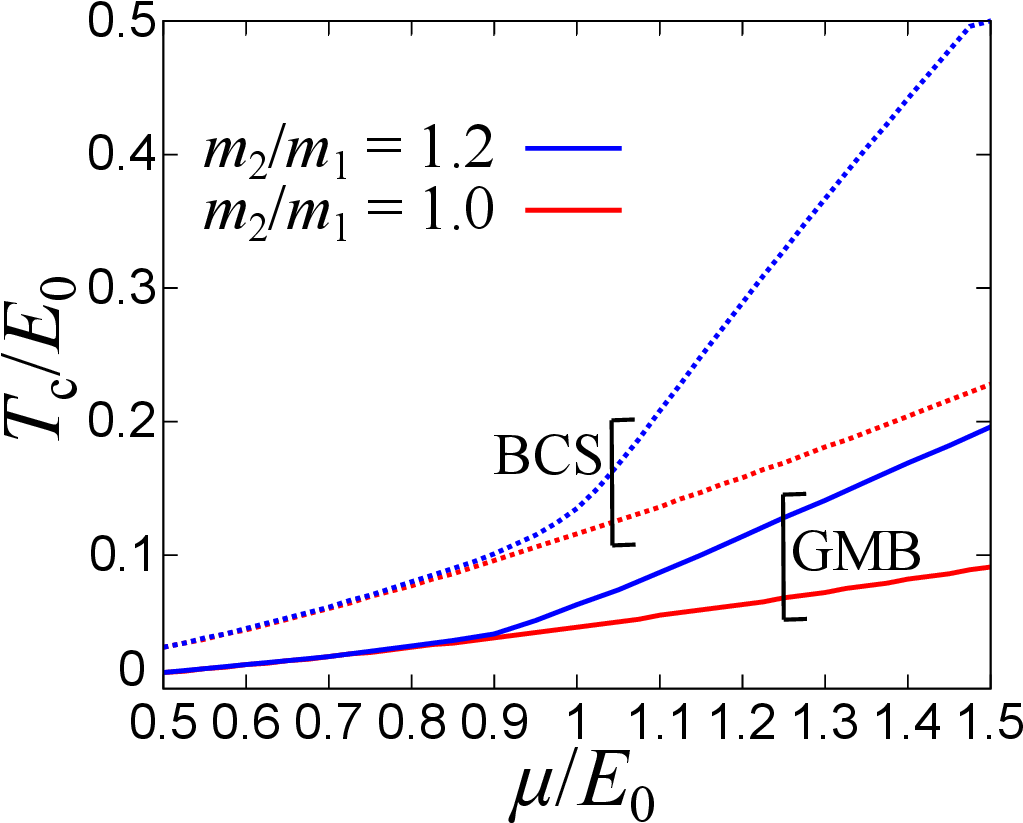}
    \caption{
    Superconducting critical temperatures $T_{\rm c}$ with the GMB correction as functions of the normalized chemical potential $\mu/E_0$ where the mass ratio is taken as $m_2/m_1=1.0$ or $m_2/m_1=1.2$ in each curve.
    For comparison, the dotted curves show the BCS result without the GMB correction.
    $\tilde{U}_{12}=10^{-3}$ and $\Lambda/k_0=10$ are used in each calculation.
    }
    \label{figs:1}
\end{figure}

Figure~\ref{figs:1} displays the superconducting critical temperature 
versus $\mu$, here for $\tilde{U}_{12}=10^{-3}$, with and without the GMB corrections.  For $m_2/m_1=1.0$ and $1.2$  
In the equal-mass case ($m_2=m_1$), $T_{\rm c}$ monotonically increases even when $\mu$ passes the bottom of the heavy band ($\mu/E_0=1$).
For comparison, we also plot the BCS result which ignores the GMB correction (dotted curve), where $T_{\rm c}$ is much larger than that with the GMB correction.  When we increase the mass ratio $m_2/m_1$ to $1.2$ in Fig.~\ref{figs:1}, 
the results exhibit a sharp increase right around $\mu/E_0=1$ 
at which band 2 starts to be occupied, in both BCS and GMB calculations, indicating the importance of the incipient heavy band.

\begin{figure}[t]
    \centering
    \includegraphics[width=7cm]{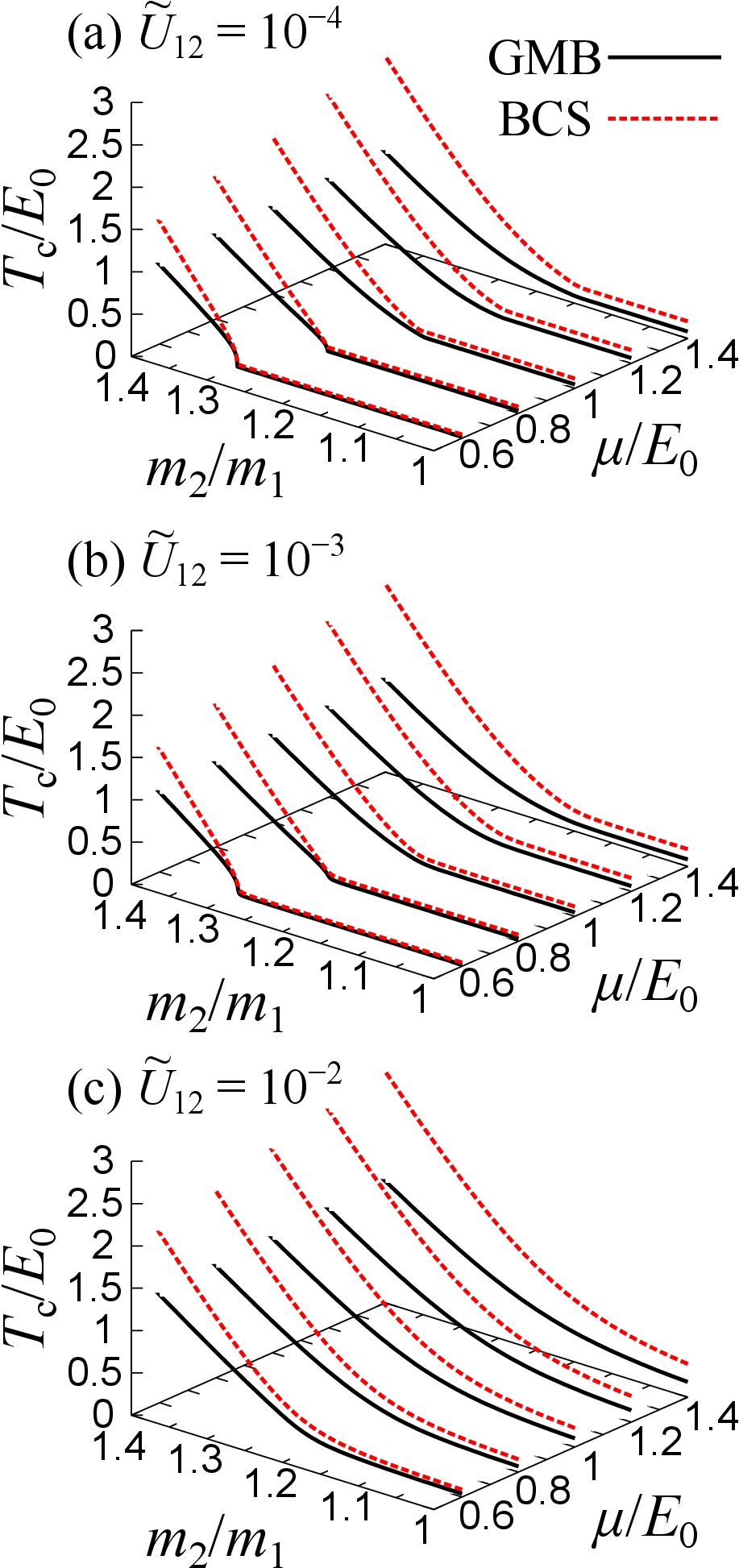}
    \caption{Superconducting critical temperature $T_{\rm c}$ in plane of the effective mass ratio $m_2/m_1$ and chemical potential $\mu/E_0$ at (a) $\tilde{U}_{12}=10^{-4}$, (b) $\tilde{U}_{12}=10^{-3}$, and (c) $\tilde{U}_{12}=10^{-2}$. The solid and dashed curves show the results with and without the GMB correction, respectively. The cutoff $\Lambda/k_0=10$ is adopted.}
    \label{fig:s2}
\end{figure}

To elaborate the dependence of $T_{\rm c}$ both on $\mu/E_0$ and $m_2/m_1$, 
we have varied the value of the pair-exchange coupling $\tilde{U}_{12}$ in Fig.~\ref{fig:s2}, where the solid (dashed) curves represent the results with (without) the GMB correction.
We can see that $T_{\rm c}$ increases with $\mu/E_0$ 
as well as with $m_2/m_1$ for the ranges of the parameters studied.
Notably, a sharp upturn in $T_{\rm c}$ for the 
mass ratio $m_2/m_1\gesim 1.2$.
This kink structure is due to the resonantly-enhanced pairing interaction inducing 
the two-body bound state in the heavy band as discussed in  Appendix A of Ref.~\cite{PhysRevResearch.4.013032}.
On the other hand, the reduction of $T_{\rm c}$ associated with the GMB correction becomes also notable in the strong-coupling regime (i.e., large $m_2/m_1$ and $\tilde{U}_{12}$) as seen in the difference between the solid and dashed curves in Fig.~\ref{fig:s2}.
Moreover, since the particle-hole fluctuations are associated with the excitations around the Fermi energy, a 
larger chemical potential leads to a stronger screening effect.
While for $\mu/E_0\leq 1$ only the particle-hole bubble $\langle\chi_{11}\rangle$ in the dispersive band dominantly contributes to the screening,
for $\mu/E_0> 1$ both of the particle-hole bubbles $\langle\chi_{11}\rangle$ and $\langle\chi_{22}\rangle$
become important because of the coexisting two Fermi surfaces from the two bands.

\begin{figure}[t]
    \centering
    \includegraphics[width=8cm]{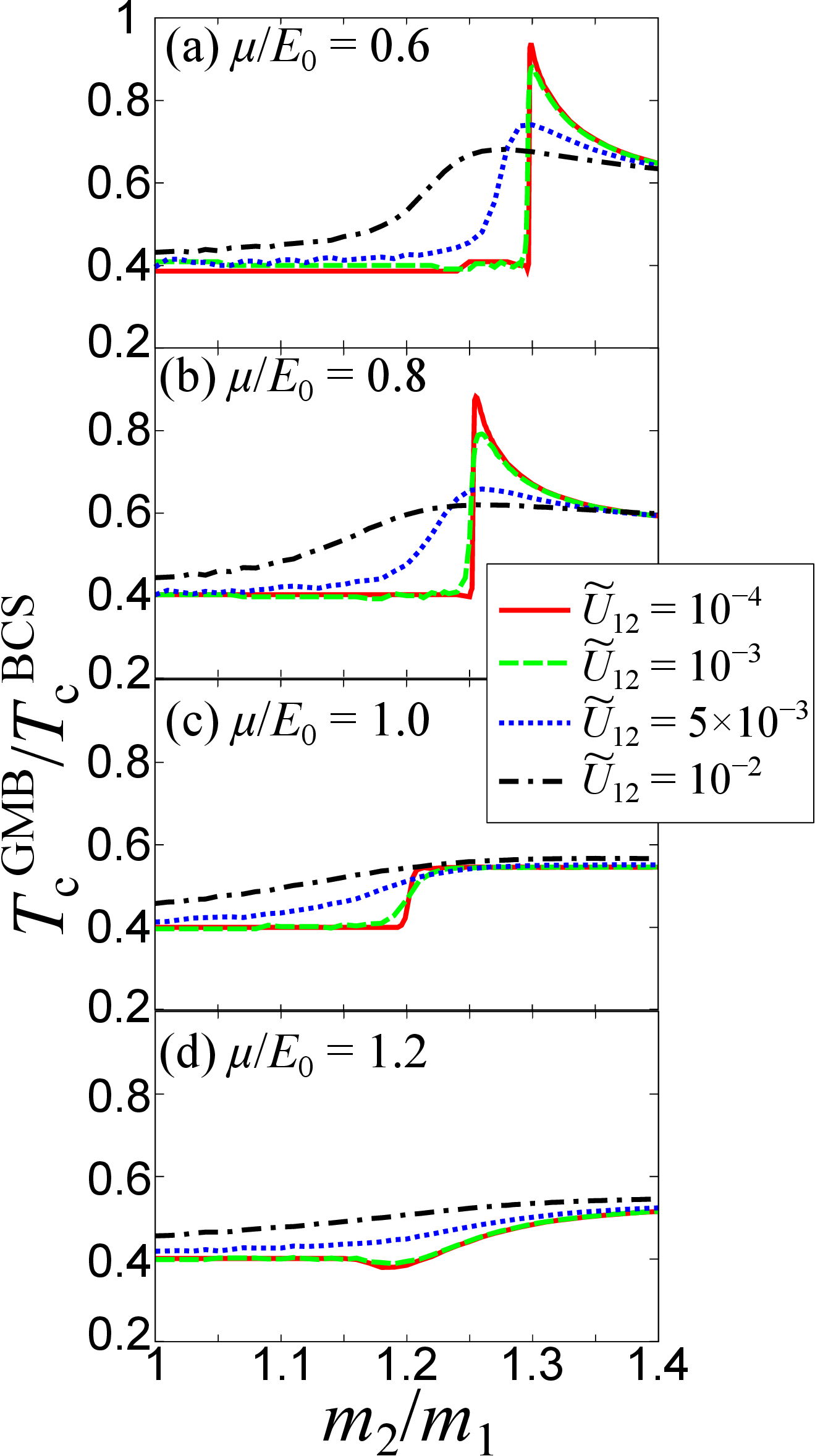}
    \caption{Ratio between $T_{\rm c}^{\rm GMB}$ and $T_{\rm c}^{\rm BCS}$ (i.e., with and without the GMB correction)
    as functions of $m_2/m_1$ at (a) $\mu/E_0=0.6$, (b) $\mu/E_0=0.8$, (c) $\mu/E_0=1.0$, and (d) $\mu/E_0=1.2$. The pair-exchange couplings are taken as $\tilde{U}_{12}=10^{-4}$ (solid), $10^{-3}$ (dashed), $5\times 10^{-3}$ (dotted), and $10^{-2}$ (dashed-dotted) in each figure. The cutoff is set to be $\Lambda/k_0=10$. }
    \label{figs:3}
\end{figure}

To quantify the GMB correction,
it is useful to examine the ratio between the superconducting critical temperatures with and without the GMB correction, denoted as $T_{\rm c}^{\rm GMB}$ and $T_{\rm c}^{\rm BCS}$, respectively.
Figure~\ref{figs:3} shows $T_{\rm c}^{\rm GMB}/T_{\rm c}^{\rm BCS}$ for $\mu/E_0=0.6-1.2$ with $\tilde{U}_{12}=10^{-4}-10^{-2}$.
As discussed in the main text, one can see the significant enhancement of $T_{\rm c}^{\rm GMB}/T_{\rm c}^{\rm BCS}$ around the Fano-Feshbach resonance at $\mu=E_0-E_{\rm b,2}/2$.
Such a behavior becomes pronounced for smaller $\tilde{U}_{12}$. 
For larger $\tilde{U}_{12}$, the peak of $T_{\rm c}^{\rm GMB}/T_{\rm c}^{\rm BCS}$ is shifted toward smaller $m_2/m_1$ and broadened as the Fano-Feshbach resonance becomes broader.
For larger $\tilde{U}_{12}$, the peak is smeared and slightly moves toward smaller $m_{2}/m_1$  
because the bound state turns into a broad resonance and the resonant energy is shifted downwards as shown in Eq.~\eqref{eq:resonance}.  


For $\mu/E_0=1.2$ in Fig.~\ref{figs:3}(d), a small dip in $T_{\rm c}^{\rm GMB}/T_{\rm c}^{\rm BCS}$ is seen.  
Such an anti-resonance-like behavior originates from the fact that both light and heavy bands tend to relatively weak-coupling  
regimes with non-negligible $\langle \chi_{11}\rangle$ and $\langle \chi_{22}\rangle$ due to the large chemical potential (i.e., a denser,  hence weak-coupling regime~\cite{PhysRevResearch.4.013032}).

\begin{figure}[t]
    \centering
    \includegraphics[width=7cm]{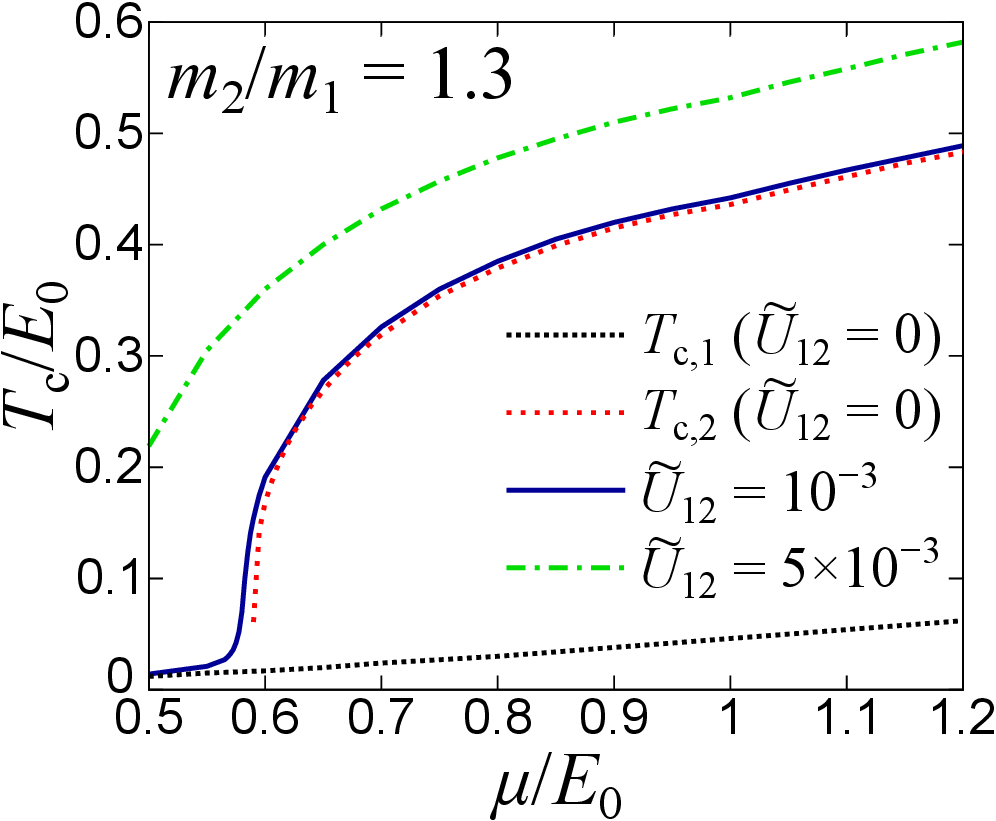}
    \caption{Calculated superconducting critical temperature $T_{\rm c}$ with the GMB approach where different interband pair-exchange couplings are taken as $\tilde{U}_{12}=0$, $10^{-3}$, and $5\times 10^{-3}$. The mass ratio $m_2/m_1=1.3$ and the cutoff $\Lambda/k_0=10$ are adopted. For $\tilde{U}_{12}=0$, we show two critical temperatures $T_{\rm c,1}$ and $T_{\rm c,2}$ obtained from \eqref{eq:tc_u12zero}.}
    \label{fig:add}
\end{figure}

\red{We note that, at $U_{12}=U_{21}=0$, the equation for $T_{\rm c}$ is decoupled into two independent equations given by
\begin{align}
\label{eq:tc_u12zero}
    1+U_{nn}^{\rm sc}\sum_{\bm{k}}\frac{1-2f(\xi_{\bm{k},n})}{2\xi_{\bm{k},n}}=0,
\end{align}
which leads to two critical temperatures $T_{\rm c,1}$ and $T_{\rm c,2}$ for each band.
Figure~\ref{fig:add} compares
$T_{\rm c,1}$ and $T_{\rm c,2}$ with   
$T_{\rm c}$ at
$\tilde{U}_{12}=10^{-3}$ and $10^{-5}$ where $m_2/m_1=1.3$ is used.
While $T_{\rm c,1}$ is small but nonzero for all the  values of $\mu$, $T_{\rm c,2}$ disappears when $\mu/E_0$ is decreased below about $0.6$ because of the vanishing population in the two-body bound state with $E_{\rm b,2}/2=0.4E_0$ in the second band (i.e., the occupation of band $2$ starts around $\mu=E_0-E_{\rm b,2}/2\equiv 0.6E_0$).
Once $U_{12}$ is introduced, we have a single $T_{\rm c}$.
With increasing $U_{12}$, $T_{\rm c}$ becomes higher than both $T_{\rm c,1}$ and $T_{\rm c,2}$ due to the Suhl-Kondo mechanism.
}

\subsection{Cutoff dependence of the superconducting critical temperature}
\label{app:2}

\begin{figure}[b]
    \centering
    \includegraphics[width=7cm]{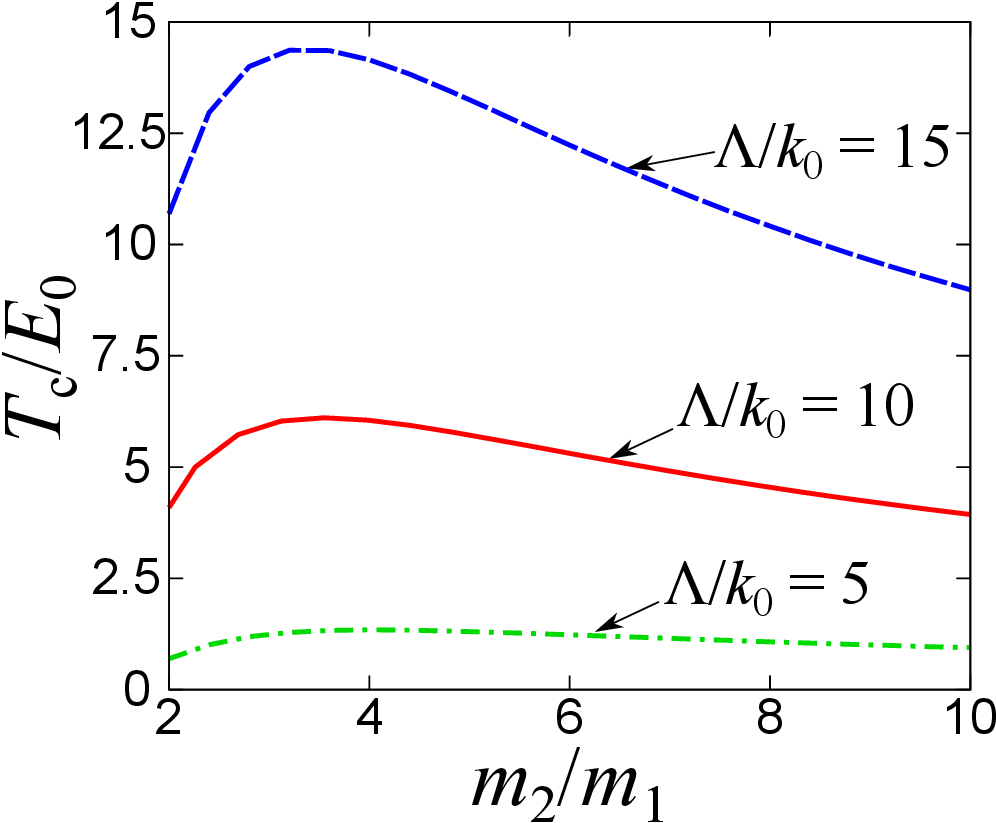}
    \caption{Calculated superconducting critical temperature $T_{\rm c}$ with the GMB approach where different cutoffs are taken as $\Lambda/k_0=5$, $10$, and $15$.
    We adopt $\mu/E_0=1$, $\tilde{U}_{12}=10^{-3}$ throughout.
    }
    \label{fig:s4}
\end{figure}

Here we examine the cutoff dependence of the GMB critical temperature $T_{\rm c}$ shown in Fig.~\ref{fig:s4}, where the cutoff is varied 
as $\Lambda/k_0=5, 10, 15$ with $\mu/E_0=1$ and $\tilde{U}_{12}=10^{-3}$.
While $T_{\rm c}$ increases for larger $\Lambda$ due to the stronger attraction originating from the flat-like second band, the GMB screening effect is seen to make 
all the curves decreasing with $m_2/m_1$ for larger $m_2/m_1$ regime. In a small $m_2/m_1$ regime, by contrast, 
$T_{\rm c}$ grows with $m_2/m_1$ for all the cases.  Thus we have a 
peaked structure persisting for all the values of $\Lambda$ studied here.   
The optimal $m_2/m_1$ that gives the peak slightly depends on the value of $\Lambda$, with the optimal $m_2/m_1$ becoming smaller 
for larger $\Lambda$ because of the saturation of the enhanced pairing effect.
Although the strong cutoff dependence is an artifact within the present continuum model, it can be interpreted as the bandwidth dependence 
if we regard the continuum model as representing band structures in lattice models.  An accurate 
analysis of the cutoff dependence will require a regularization for the cutoff in a renormalization scheme in the continuum model.
We have numerically confirmed that the peaked behavior of $T_{\rm c}$ persists even for larger $\Lambda$ with different values of $\mu$.

\end{document}